\documentclass[12pt]{article}
\usepackage{times}
\usepackage{float}
\usepackage{mathptmx}
\usepackage[utf8]{inputenc}
\usepackage{amsmath}
\usepackage{amssymb}
\usepackage{cite}
\usepackage{caption}
\usepackage{subcaption}
\usepackage{latexsym}
\usepackage{graphicx}
\usepackage{epstopdf}
\usepackage{wrapfig}
\usepackage{enumerate}
\usepackage{graphicx}
\usepackage{wrapfig}
\usepackage{mdwlist}
\usepackage{fancyhdr}
\usepackage{fixltx2e}
\usepackage{color}
\usepackage{txfonts}
\DeclareMathOperator{\sech}{sech}
\DeclareMathOperator{\csch}{cosech}

\begin{document}
	\begin{center}
		\vspace{0.4cm}{\large \bf \textsc{ Viscous Ricci Dark Energy Cosmological Models in Brans-Dicke Theory }}\\
		\vspace{0.4cm}{ M.Vijayasanthi$^{1,*}$, T.Chinnappalanaidu$^{1,2}$, S. Srivani Madhu$^{1}$}\\
		$^{1}$Department of Applied Mathematics, Andhra University, Visakhapatnam 530003, India\\
		$^{2}$Department of Mathematics, Vignan's Institute of Information Technology(Autonomous), Visakhapatnam 530049, India\\
		{$^*$gv.santhi@live.com}
		\end{center}
	\begin{abstract}
The whole article deals with the analysis of the cosmic model of Ruban's space-time in the context of a bulk viscosity impact in the form of Ricci dark energy within the framework Brans-Dicke theory (Brans and Dicke, \textit{Phys. Rev.} \textbf{124}, 925 (1961)). We believe that outer space is filled with dark matter and viscous Ricci dark energy (VRDE) under the pressureless situation. The velocity and rate at which the Universe is expanding are presumed to be proportional to the coefficient of total bulk viscosity, is in the form, $\xi_0+\xi_1 \frac{\dot{a}}{a} +\xi_2 \frac{\ddot{a}}{\dot{a}}$, where $\xi_0$, $\xi_1$ and $\xi_2$ are the constants. To solve the RDE model's field equations, we utilize the relation among the metric potentials and also the power-law relation among the average scale factor $a(t)$ and scalar field $(\phi)$. To examine the evolutionary dynamics of the Universe, we investigate the deceleration parameter$(q)$, jerk parameter$(j)$, EoS parameter $(\omega_{de})$, Om$(z)$, stability of the obtained models through the square speed of the sound $(v_s^2)$, $\omega_{de}-\omega_{de}'$ plane, statefinder parameter planes $(r,s)$ \& $(q,r)$ and presented via graphical representation. By the end of the discussion, VRDE model was found to be compatible with the present accelerated expansion of the Universe.
\end{abstract}
\section{Introduction}
 \hspace*{0.6 cm} One of the most fundamental questions in modern theoretical cosmology is, whether the genesis of the Universe was singular or non-singular. This question is equivalent in asking if the Big Bang theory or the Big Bounce theory actually describes the evolution of our Universe. Naturally thinking, the initial singularity described by the Big Bang theory, is a mentally more convenient description, since we can easily imagine a zero sized Universe, with infinite temperature and energy density, and also in which all fundamental interactions are unified under the yet unknown same theoretical framework. Moreover, no one can actually exclude a cyclic cosmological evolution, in which the Universe never shrinks to zero \cite{1q2,1q3}. However, Modern advanced cosmology is assumed to have begun in 1917 with Albert Einstein's publication of his last correction to General Relativity(GR) in his article ``cosmological considerations of the general theory of relativity" \cite{ein}. Even over a century, this theory still exists as a valid theory with modifications and exhibits a prominent role in the study of the cosmos. However, for over years, various experiments have unveiled intriguing new secrets of the Universe, looking deep into space with unprecedented clarity and uncovering the latest advancements in studies about the cosmos \cite{agr,sp,spe}, finding many astonishing facts from the measurement of Supernovae type Ia accompanied by Cosmic Microwave Background Radiation (CMBR) and the large structures \cite{clb,mte} substantiating that the Universe is in its accelerated expansion state. Two essential components that drive this expansion and the characterization of the cosmos are dark matter (DM) and dark energy (DE). We can pronounce that, DM, which covers about 26\% of the total matter density, is accountable for the structure formation and the clustering of galaxies \cite{js,lb,cs,mv}, whereas DE forming 69\% of the cosmic energy with negative pressure is responsible for the accelerated expansion of the Universe and the remaining part of the cosmos is related to the baryonic (visible) matter.
 
  Despite the awareness of the remarkable success of standard cosmology, it is yet unable to resolve the significant issues that include finding the best DE candidate. There are two ways to describe dark energy models. One is a fluid description (Nojiri et al. \cite{2005b}; Nojiri and Odintsov \cite{2005a}; Stefancic \cite{ste2005}) and the other is to describe the action of a scalar field theory. In the former fluid description, we express the pressure as a function of $\rho$ (in more general, and other background quantities such as the Hubble parameter H). On the other hand, in the latter scalar field theory we derive the expressions of the energy density and pressure of the scalar field from the action.  The basic candidate for DE can be the cosmological constant $\Lambda$, which is associated with the energy vacuum with stable pressure and energy that can be distinguished by the (EoS) $\omega_\Lambda=-1$ \cite{sw}. The various cosmological observational data supports the $\Lambda$ cold dark matter ($\Lambda$CDM) model, in which the cosmological constant $\Lambda$ plays a role of dark energy in general relativity. At the current stage, the $\Lambda$CDM model is considered to be a standard cosmological model. However, the theoretical origin of the cosmological constant $\Lambda$ has not been understood yet \cite{sw}. However, two frequent issues ``fine-tuning" and ``cosmic coincidence" are being faced. Here, the fine-tuning issues call attention to the variation among the theoretical and the observable value of vacuum or cosmological energy constant, whereas the hassle of coincidence is the co-occurrence of DM and DE. A number of models for dark energy to explain the late-time cosmic acceleration without the cosmological constant has been discussed by Bamba et al. \cite{bamba1q1}. To reduce the persistent cosmological problem and justify the accelerated expansion, some alternative models are suggested by either modifying the right side of field equations of Einstein by considering the specific energy-momentum tensor $T_{\mu \nu}$ forms, which includes quintessence \cite{smc}, K-essence\cite{tch}, Chaplygin gas model \cite{aka}, holographic DE\cite{gt,lsu}, new agegraphic dark energy \cite{hwe} etc., or modifying the left side of field equations of Einstein, we derive the modified gravity that include $f(R)$ gravity \cite{sco} and scalar tensor theories \cite{lme}. The pioneering research on scalar-tensor theories has been carried out by Brans and Dicke \cite{cbr} to include Mach’s principle into gravity which is known as Brans-Dicke theory (BDT).
 
A pragmatic theory of gravity, BDT is predominantly a gravitational theory in which its field is described by the tensor field, which is ascertained by the distribution of mass energy in the Universe and replaces the gravitational constant. This theory has initially put forward in the 1960s by Brans and Dicke \cite{cbr} as a variant of GR. Contrary to GR, BDT can indeed satisfy all current gravitational experiments and also illustrate intuitively, without the addition of DE to the expansion of the Universe \cite{cmw,obe,loi,ybi}. Obviously, as the parameter $\omega\rightarrow\infty$, GR is retrieved from BDT which summons the BDT as universality to GR. Hence, to assure its free fall generality (Equivalence principle), an uncomplicated modification to GR is made maintaining a pure metric relation of matter and gravity \cite{xll}. The planetary system observational measure \cite{bbe} and the phenomenon of accelerated expansion of the cosmos \cite{cma} certainly accounted for replacing the gravitational constant with $\phi=\frac{1}{8\pi G}$ and connecting $\phi(t)$ to gravity with a constant $\omega$. Also, CMB and large scale observational data \cite{vac,sts,fwu,fwux} uphold this theory making it stronger than GR with a dynamical framework, that evokes huge curiosity in present-time cosmology. Prasad et al.\cite{bd1} have studied constraining Bianchi type V Universe with recent $H(z)$ and BAO observations in BDT of gravitation. Hatkar et al. \cite{bd2} have worked on viscous holographic dark energy in BDT of gravitation. Shaikh \cite{bd3} has studied viscous dark energy cosmological models in BDT of gravitation. Koyama \cite{bd4} has studied testing BDT gravity with screening by scalar gravitational wave memory. Sharif and Majid \cite{bd5} have studied extended gravitational decoupled solutions in self-interacting BDT. Singh and Soibam \cite{bd6} have studied anisotropic models with generalized hybrid expansion in BDT of gravity. Hou \cite{bd7} has gravitational memory effects in BDT. Roy et al., \cite{bd8} have discussed some characteristics of the accelerated expansion of the Universe in the framework of BDT. Sharif and Majid \cite{bd9} have studied the effects of charge on decoupled solutions in self-interacting BDT. Tahura et al. \cite{bd10} have investigated the connection between gravitational-wave memory effects, asymptotic symmetries, and conserved quantities in Brans-Dicke theory and computed the field equations in Bondi coordinates to define a set of boundary conditions representing the asymptotically flat solutions. Santhi and Babu \cite{ysb} have studied axially symmetric VRDE in BDT of gravitation. Santhi et al. \cite{newtcn1} have investigated the some Bianchi type viscous holographic dark energy cosmological models in the BDT of gravitation. BDT has been explored by many cosmologists in the framework of DE \cite{msri,cps,xga}.   
 
However, the DE has always been a mysterious component in the study of the Universe. A thought has been given for considering DE as a consequence of quantum gravity, such that, the core idea of quantum gravity- ``the Holographic Principle" becomes an essential part of resolving the DE issue. In recent years, there has been a significant attempt to reconcile vacuum energy density to the holographic theory of quantum gravity. This en routes to a proposal of a new DE model: the Holographic dark energy (HDE) \cite{sdh}, on the basis of the holographic principle of quantum field theory\cite{gt,lsu,rbo}, relating the ultra-violet (UV) cutoff $\Lambda$ to the infrared (IR) cutoff $L$. Assuming the Hubble length as IR-cutoff, Li \cite{mli} has suggested HDE as $\rho_{\Lambda}=3c^2M_p^2 L^{-2}$, taking $L$ as the IR-cutoff, $M_p= (8\pi G)^{-1/2}$ as the Planck mass on a reduced scale and $c^2$ as a definite constant. As discussed earlier, the fine tuning issue is also solved by the HDE model as suggested by Li \cite{mli}, but creates an incorrect EoS of DE. The IR interface correspondence with the Universe's large-scale structure and the UV interface is correlated to the energy of the vacuum. A wide range of HDE models are proposed recently based on the IR-cutoff options, such as event horizon. IR-cutoff in terms of event horizon has been considered by Li et al. \cite{mlix} to study the Universe's expansion. The corresponding model, however, faces challenges of causality violation.

Subsequently, Gao et al. \cite{cgaof} implemented an HDE model where they reversed the Ricci scalar curvature to replace the future event horizon and named it the ``RDE model''. The energy density for RDE is given by
\begin{equation}\label{eq1}
	\rho_{de}=3 \alpha (\dot{H}+2H^2),
\end{equation} 
where $\alpha$ and $H$ are the density parameter (without dimensions) and the Hubble parameter respectively. It has been learned that this model overcomes the causality problem and addresses the coincidence problem.  A phantom-based consolidating perspective to the early and late Universe cosmology was put forward by Nojiri and Odintsov \cite{snoj}. Further, the authors have proposed a generalized HDE wherein infrared cutoff is determined by combining the FRW parameters: Hubble constant, particle and future horizons, cosmological constant, and Universe lifetime (if finite). Gao et al.\cite{cgaof} have suggested the use of Ricci DE that can be interpreted as a form of holographic DE, with the reciprocal of Ricci scalar square root as its infrared cut-off. Whenever the vacuum density emerges as a parameter of independently stored energy, it may be consistent with present astrophysical findings. The VRDE model was discussed by Feng and Li\cite{cjf}, who believed that the linear barotropic fluid and RDE are bulk viscosity. They have also discussed about the outcome of some principles on cosmic evolution, the RDE model with a more general form of bulk viscosity. Dixit et al. \cite{hr2} have studied a model for modified holographic RDE in the gravitation theory of BDT. Chakrabarti et al. \cite{vhr3} have studied the bulk viscous pressure in scalar fields and holographic RDE considered in the modified gravity framework. Kumar and Singh \cite{vr4,vr5,vr8,vr9} have studied VRDE model with matter creation, exact solution and observational tests; the generalized second law of thermodynamics in VRDE; RDE model with bulk viscosity and have also studied observational constraints on VRDE model. Santhi and Babu \cite{vr7} have studied Kantowski–Sachs VRDE model in Saez–Ballester theory of gravitation.  

As revealed by the current investigations, the transition phase has a vital part in explaining the growth of the space. We have studied viscous models and observed that they help in understanding the phase transition and hence can be a possible candidate for DE. Some of the authors \cite{vr8,sasi} had a detailed study regarding the effects of bulk viscosity in HDE model. According to thermodynamics \cite{eck,lan,isra}, it is observed that viscosity has a role to play as DE in our Universe. A constant bulk viscosity or otherwise called constant DE along with dark cold matter can be considered one of the simplest principles of the kind. A widely investigated way is taking the bulk viscosity as a Hubble parameter with a linear function, that is closely aligned with the observed late-time acceleration. Our article focuses on RDE model along bulk viscosity, aspiring from the research of Feng and Li\cite{cjf}, Singh and Kumar \cite{vr8} and Cataldo et al. \cite{cata}. By extending Singh and Kumar's work with a generalized form of bulk viscosity coefficient, which is further examined analytically and performs reasonably well when compared to observational data. Many authors\cite{jdb,ibr,djl,ibre,ibrevi,newtcn2} have researched homogeneous and non-homogeneous viscous cosmology. The Universe evolution through a cosmological model with bulk viscosity was studied by Ren and Meng\cite{jr}. Discussion of the phase transition of the viscous early Universe was done by Tawfik and Harko \cite{ata}. Singh and Kumar \cite{cpsi,cpsin} have inspected different features of viscosity in $f(R,T)$ gravity taking Hubble horizon as IR cutoff. Nojiri and Odintsov \cite{3q7} have introduced the DE Universe EoS with inhomogeneous, Hubble parameter dependent term that comes from time dependent viscosity considerations and modification of general relativity. Capozziello et al. \cite{3q8} have investigated the effects of viscosity terms depending on the Hubble parameter and its derivatives in the DE equation of state. Various explanations can be used to justify the analysis of non-perfect fluids. In spite of various macroscopic physical systems, such as the large-scale structure of matter and radiation of the cosmos, which resembles perfect fluids, we could not eliminate the various components of DE, whose origin is not known.

For a number of reasons, in a classical cosmological scenario, an inhomogeneous exact solution of Einstein's field equations for an imperfect fluid is necessary to be obtained; among them, one is connected to the existing entropy of the Universe. It is already recognized that, the rate of entropy generated from a non-adiabatic procedure in an originally homogeneous backdrop appears to be inadequate to comprehend the high entropy per particle of the Universe \cite{in1,in2}. By assuming the spatial homogeneity, which is believed to be reasonable in an averaged sense, cosmological elucidation to Einstein's field equations is broadly determined \cite{in3}. However, it turns out to be inappropriate on galactic and smaller scales; hence, a broad category of inhomogeneous models is compared with homogeneous ones. To be fully addressed, any detailed concerns, to give an example like, star system origination or the intricate composition of black-body radiation, should finally be directed to such inhomogeneous models.

Ever since Szekers learned that the Universe is filled up with extraterrestrial or cosmic dust \cite{in4}, much concentration has been given to inhomogeneous cosmic models \cite{in5,in6,in7,in8,in9,in10}. Also, in recent times, the Szekers space-time is expanded by proposing a recent radioactive constituent and an electromagnetic field \cite{in5}. Tomimura and Waga \cite{nto} have observed that whenever space-time gains symmetry, self-consistent solutions are possible if an electrostatic field is added as a source term for the Szekers metric of class II, reducing the Szekers metric to the inhomogeneous basic form as initially contemplated by Ruban \cite{var}. In this article, we establish the exact solutions in the constitution of Rubans metric. Lima and Nobre \cite{ja} have studied the electromagnetic field model in Rubans's Universe. Also, the thermodynamics of the Ruban’s Universe and inhomogeneous two cosmological models is investigated by Lima and Tiomno \cite{jasl,jas}.  Tomimura and Waga \cite{nto} have derived reliable solutions for Ruban’s space-time with dust, radiation, and electromagnetic field and Mete et al. \cite{vgm} investigate the cosmological model of Ruban’s with a significant source of stress in the general relativity. The energy-momentum distribution of the Ruban’s Universe in GR and teleparallel gravity is studied by Aktas \cite{cak}. Very recently, Santhi and Naidu \cite{tcn2} have studied Renyi HDE in a scalar tensor theory with Ruban's Universe.

Taking some inspiration from the above mentioned explorations, we have engrossed our research on Ruban's cosmological model with VRDE in BDT of gravitation. This article is further studied in the following sections. Section 2 deals with the derivation of BD field equations for Ruban's line element in the presence of RDE. In section 3, the solutions for these field equations for $\kappa=0,1 \& -1$ are found. Section 4 is devoted to the physical parameters of our model, and we conclude our article by summarizing the results in the final section.
  
\section{Metric and field equations}
In an approach to understanding the structure of the Universe, here we consider Ruban's space-time \cite{var} whose metric takes the following form:
\begin{equation}\label{eq2}
ds^2=dt^2-Q^{2}(x,t) dx^2-R^{2}(t)(dy^2+h^2dz^2),
\end{equation}
\begin{equation}\label{eq3}
\text{where}\quad h(y)=\frac{\sin\sqrt{\kappa}y}{\sqrt{\kappa}}= \left\{\begin{array}{c} \sin y ~~~~if~~~~ \kappa=1 \\ y ~~~~if~~~~ \kappa=0\\ \sinh y  ~~~~if~~~~ \kappa=-1  \end{array}\right.
\end{equation}
and $\kappa$ shows the curvature parameter of the homogeneous $2$-spaces $x$ and $t$. Opting $Q^2(x,t)=Q^2(t)$, in \eqref{eq2}, LRS Bianchi type--I, III and Kantowski-Sachs space times are obtained. This model is a special form of Szekers Universe \cite{ja}, which represents the inhomogeneous and anisotropic Universe. The volume $(V)$, average scale factor $(a(t))$ and the Hubble parameter($H$) of the Ruban's space-time are defined as 
\begin{align}\label{eq4a}
V =& [a(t)]^3 = QR^2h, \\&  \& \quad H=\frac{\dot{a}}{a}\label{eq5a}.
\end{align}
The action of BDT in the presence of matter with Lagrangian $L_m$ in its canonical form is given by 
\begin{equation}\label{eq4}
	\mathcal{S}= \int d^4x\sqrt{-g}\left(-\phi \mathfrak{R}+ \frac{\omega}{\phi} \triangledown_\mu \phi \triangledown_\mu \phi+L_m \right),
\end{equation}
where $\phi$ is the BDT scalar field depicting the Newton constant's inverse, which is permissible to shift with space-time, $\mathfrak{R}$ is the scalar curvature, $\omega$ is the BD constant. Varying the action in Eq.\eqref{eq4} w.r.t. the metric tensor $g_{\mu \nu}$ and the scalar field $\phi$ we procure the field equations as:
\begin{equation}\label{eq5}
G_{\mu \nu}=-8\pi \phi^{-1} T_{\mu \nu}-\omega \phi^{-2} \left(\phi_{,\mu}\phi_{,\nu}-\frac{1}{2}g_{\mu \nu} \phi_{,\kappa}\phi^{,\kappa}\right)-\phi^{-1} \left(\phi_{\mu;\nu}-g_{\mu \nu} \phi_{;\kappa} ^{,\kappa}\right)
\end{equation} 
and 
\begin{equation}\label{eq6}
	\phi_{;\kappa} ^{,\kappa}=8\pi\left(3+2\omega\right)^{-1}T,
\end{equation} 
where $G_{\mu \nu}=\mathfrak{R}_{\mu \nu}-\frac{1}{2}\mathfrak{R}g_{\mu \nu}$ is an Einstein tensor, $T_{\mu \nu}$ is the stress energy tensor of the matter and $\omega$ is the dimensionless constant.\\
The energy conservation equation is given as
\begin{equation}\label{eq7}
	T^{\mu \nu}_{;\nu}=0,
\end{equation} 
is a outcome of Eqs. \eqref{eq5} and \eqref{eq6}.

\noindent For a viscous fluid, the energy momentum tensor \cite{wz} is considered as
\begin{equation}\label{eq8}
T_{\mu \nu}=(\rho_{m}+\rho_{de})u_{\mu}u_{\nu}+\overline{p}_{de}(g_{\mu \nu}+u_{\mu}u_{\nu}),
\end{equation}  
where $\rho_{de}$ and $\rho_{m}$ are the energy densities of VRDE and DM, respectively. In Eckart's \cite{ce} first order thermodynamics, the effective pressure ($\overline{p}_{de}$) of VRDE is given by 
\begin{equation}\label{eq9}
	\overline{p}_{de}=p_{de}-3\xi H,
\end{equation} 
where $H$ is the Hubble parameter and $\xi$ is the bulk viscous coefficient.

From Eq. \eqref{eq8}, the field Eqs. \eqref{eq5} and \eqref{eq6} for the line element in Eq. \eqref{eq2} are as follows:
\begin{align}\label{eq10}
	\frac{2\ddot{R}}{R}+\frac{\dot{R}^2}{R^2}+\frac{\kappa}{R^2}+\frac{\omega}{2}\frac{\dot{\phi}^2}{\phi^2}+\frac{2\dot{R}\dot{\phi}}{R\phi}+\frac{\ddot{\phi}}{\phi}=&-\frac{8\pi\big(p_{de}-3\xi H\big)}{\phi},\\
\label{eq11}
	\frac{\ddot{Q}}{Q}+\frac{\dot{R}\dot{Q}}{RQ}+\frac{\ddot{R}}{R}+\frac{\omega}{2}\frac{\dot{\phi}^2}{\phi^2}+\frac{\ddot{\phi}}{\phi}\left(\frac{\dot{Q}}{Q}+\frac{R}{R}\right)+\frac{\ddot{\phi}}{\phi}=&-\frac{8\pi\big(p_{de}-3\xi H\big)}{\phi},\\
\label{eq12}
	\frac{2\dot{R}\dot{Q}}{RQ}+\frac{\dot{R}^2}{R^2}+\frac{\kappa}{R^2}-\frac{\omega}{2}\frac{\dot{\phi}^2}{\phi^2}+\frac{\dot{\phi}}{\phi}\left(\frac{\dot{Q}}{Q}+\frac{2\dot{R}}{R}\right)=&\frac{8\pi(\rho_{m}+\rho_{de})}{\phi},
\\
\label{eq13}
\&~~~	\dot{\phi}\left(\frac{\dot{Q}}{Q}+\frac{2\dot{R}}{R}\right)+\ddot{\phi}=\frac{8\pi}{3+2\omega}\bigg(\rho_{m}+\rho_{de}&-3\big(p_{de}-3\xi H\big)\bigg).
\end{align}
Moreover, the energy conservation equation gives:
\begin{equation}\label{eq14}
\dot{\rho}_{m}+\dot{\rho}_{de}+3\left(\frac{\dot{Q}}{Q}+\frac{2\dot{R}}{R}\right)\left(\rho_{m}+\rho_{de}+\big(p_{de}-3\xi H\big)\right)=0.
\end{equation}
Here, the overhead dot denotes differentiation with respect to $t$ and $\kappa=0,1\,\, \& -1$.
\section{Solutions of the field equations}
The system of four independent field equations \eqref{eq10}-\eqref{eq13} with seven unknowns namely $R$, $Q$, $p_{de}$, $\xi$, $\rho_{m}$, $\rho_{de}$ \& $\phi$. To get a deterministic solution, we choose three plausible physical conditions.
\begin{itemize}
	\item [(i)] The shear scalar $\sigma$ proportional to scalar expansion $\theta$ \cite{cbc,throne}, giving a relationship between the metric potentials as
	\begin{equation}\label{eq15}
	Q=(xR)^m,	
	\end{equation}
	where $m$ is a positive constant \& $m\neq 1$ $\left(i.e., 0<m<1\,\, \text{or}\,\, \, m>1 \right)$.
	\item [(ii)] The scalar field $\phi$ as a function of average scale factor `$a$' \cite{sunil}, $i.e.,$
	\begin{equation}\label{eq16}
	\phi=\phi_0 a^n, \quad \text{where}\quad \phi_0,\: \& \:n \quad \text{are constants}.
	\end{equation} 
	\item [(iii)] We assume a parameterized bulk viscosity such as \cite{jr}
	\begin{equation}\label{eq16a}
	\xi=\xi_0+\xi_1 \frac{\dot{a}}{a} +\xi_2 \frac{\ddot{a}}{\dot{a}}, 
	\end{equation}
	where $\xi_0$, $\xi_1$ and $\xi_2$ are the constants.
\end{itemize}

Now from \eqref{eq10} \& \eqref{eq11} and \eqref{eq15} \& \eqref{eq16} we derive the below equation
\begin{equation}\label{eq17}
\frac{\ddot{R}}{R}+\Bigg(\frac{3+3m+nm+2n}{3}\Bigg)\frac{\dot{R}^2}{R^2} +\frac{\kappa}{R^2(1-m)}=0.	
\end{equation}
Three probable scenarios for $\kappa=0, 1$ \& $-1$, when substituted in Eq. \eqref{eq17} are obtained and discussed in the following subsections.
\subsection{Model--I}
If $\kappa=0$, then by solving Eq. \eqref{eq17}, we get 
\begin{equation}\label{eq18}
R=\big(\gamma\big(c_1t+c_2\big)\big)^{\frac{1}{\gamma}},
\end{equation}
where $c_1$ and $c_2$ are integration constants and $\gamma=\frac{9+6m+2nm+4n}{3}$.\\
From Eqs. \eqref{eq15} and \eqref{eq18}, we have
\begin{equation}\label{eq19}
Q=x^m \big(\gamma\big(c_1t+c_2\big)\big)^{\frac{m}{\gamma}},
\end{equation}
and the scalar field $\phi$ is given by 
\begin{equation}\label{eq20}
\phi=\phi_0\bigg(x^{\frac{m}{3}} y^{\frac{1}{3}} \big(\gamma\big(tc_{1}+c_{2}\big) \big)^{\frac{2+m}{\gamma}}\bigg)^{n}.
\end{equation}
Hence, from the above expressions of $R$ \& $Q$, the metric \eqref{eq2} takes the form
\begin{equation}\label{eq21}
ds^2=dt^2-x^{2m} \big(\gamma\big(c_1t+c_2\big)\big)^{\frac{2m}{\gamma}} dx^2-\big(\gamma\big(c_1t+c_2\big)\big)^{\frac{2}{\gamma}}(dy^2+y^2dz^2).
\end{equation}
The volume $(V)$, average scale factor $(a)$ and the Hubble parameter ($H$) are obtained as 
\begin{align}\label{eq22}
V=&x^m y (\gamma \big(tc_1+c_2\big))^{\frac{m+2}{\gamma}}, \\ \label{eq23}
a=&x^{\frac{m}{3}}y^\frac{1}{3} (\gamma \big(tc_1+c_2\big) )^{\frac{1}{3}\frac{2+m}{\gamma}}, \\
\label{eq24}
\& \quad H=&\frac{c_1(2+m)}{3\gamma\big(tc_1+c_2\big)}.
\end{align}
From Eqs.\eqref{eq8} and \eqref{eq24}, we get the energy density of VRDE as
\begin{equation}\label{eq26}
\rho_{de}= \frac{1}{3}\bigg(\frac{\alpha c_1^2(2+m)(4+2m-3\gamma)}{\gamma^2\big(\big(tc_1+c_2\big)\big)^2}\bigg).
\end{equation}
The energy density of matter is
\begin{equation}\label{eq27}
\rho_{m}=\frac{\varPhi_{1}}{144\pi\gamma^2\big(tc_1+c_2\big)^2},
\end{equation}
$$\left.
\begin{aligned}
\text{where}\quad \varPhi_{1}=&\bigg(\bigg(\phi_0((6n-n^2\varw)m^2-(4n^2\varw-24n-36)m\\&-4\varw n^2+24n+18)\big(x^{\frac{m}{3}}y^\frac{1}{3}\big(\gamma\big(tc_{1}+c_{2}\big) \big)^{\frac{2+m}{3\gamma}}\big)^n-96\big(m-\frac{3}{2}\gamma+2\big)\pi(2+m)\alpha\bigg)c_1^2\bigg).
\end{aligned}\right\}
$$
The effective pressure is obtained as
\begin{equation}\label{eq28}
\overline{p}_{de}=\frac{\bigg(\bigg(x^\frac{m}{3}y^\frac{1}{3}\big(\gamma\big(tc_{1}+c_{2}\big) \big)^{\frac{2+m}{3\gamma}}
	\bigg)^n c_1^2\phi_0 \bigg(6(\gamma-2)(m+2)n+36\gamma-(m+2)^{2} (\varw+2)n^{2}-54 \bigg)\bigg)}{\pi\gamma^2\big(\big(tc_1+c_2\big)\big)^2},
\end{equation}
The viscosity coefficient is 
\begin{equation}\label{eq29}
\xi=\xi_{0}+\frac{\xi_{1}(2+m)c_1}{3\gamma\big(tc_1+c_2\big)}-\frac{\xi_{2}c_1^2(2+m)}{3\gamma\big(tc_1+c_2\big)^2}.
\end{equation}
The proper pressure is obtained as 
\begin{equation}\label{eq30}
p_{de}=\frac{\varPhi_{2}}{3\pi\big(tc_1+c_2\big)^3\gamma^2},
\end{equation}
$$
\left.
\begin{aligned}
\text{where}\quad \varPhi_{2}=&\bigg(\bigg(-\frac{1}{48}\phi_0\big(tc_1+c_2\big)\big(n^2(\varw+2)m^2+4\big((\varw+2)n-\frac{3}{2}\gamma+3\big)nm+(4\varw+8)n^2+(-12\gamma+24)n\\&-36\gamma+54\big)c_1\big(x^\frac{m}{3}y^\frac{1}{3} \big(\gamma\big(tc_1+c_2\big)\big)^{\frac{(2+m)}{3\gamma}}\big)^n+\bigg((3t^2\gamma \xi_{0}+\xi_{1}(2+m)t-\xi_{2}(2+m))c_1^2+(6t\gamma \xi_{0}\\&+\xi_{1}(2+m))c_2c_1+3c_2^2\gamma \xi_{0}\bigg)
\pi(2+m)\bigg)c_1\bigg).
\end{aligned}\right\}
$$
\begin{figure}[H]
	\centering
	\begin{minipage}{0.50\textwidth}
		\includegraphics[width=1\linewidth]{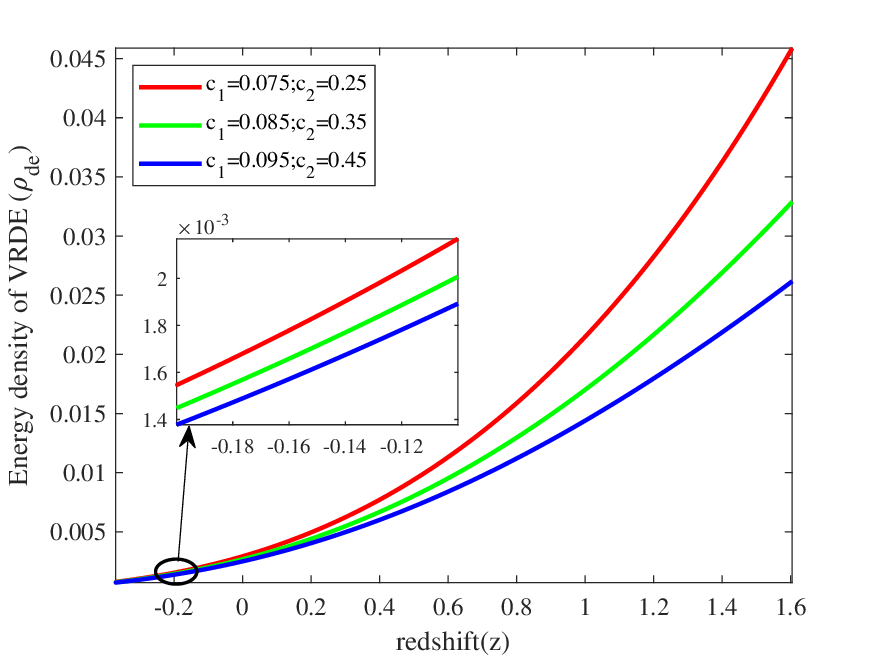}
		\caption{ Plot of  energy density of VRDE \\$(\rho_{de})$ versus redshift $(z)$.}
		\label{fig:rhode0}
	\end{minipage}
	\begin{minipage}{0.50\textwidth}
		\includegraphics[width=1\linewidth]{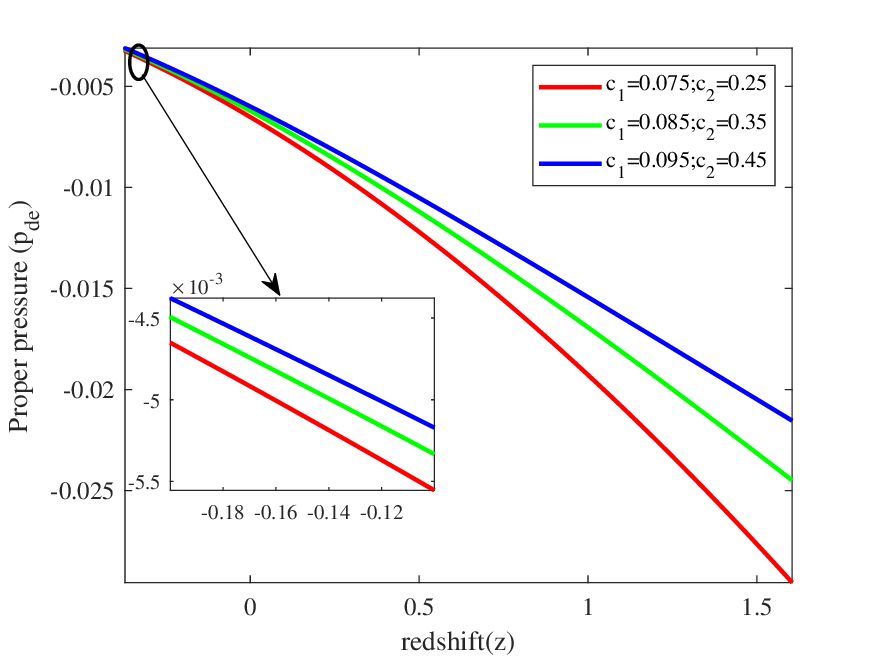}
		\caption{ Plot of proper pressure$(p_{de})$ versus redshift $(z)$.}
		\label{fig:pp0}
	\end{minipage}
\end{figure}
\subsection{Model--II}
If $\kappa=1$, then the Eq. \eqref{eq17} for $n=-3$ yields, 
\begin{equation}\label{eq31}
R=\Bigg(\frac{k_{2}}{k_{1}} \sinh\big(k_{1}t+k_{3}\big)\Bigg).
\end{equation}
From the Eqs. \eqref{eq15} and \eqref{eq31}, we have
\begin{equation}\label{eq32}
Q=x^{m}\Bigg(\frac{k_{2}}{k_{1}} \sinh\big(k_{1}t+k_{3}\big)\Bigg)^{m},
\end{equation}
and the scalar field $\phi$ is given by 
\begin{equation}\label{eq33}
\phi=\phi_{0}x^{-m} \big(\sin y\big)^{-1} \bigg(\frac{k_{2}\sinh\big(k_{1}t+k_{3}\big)}{k_{1}}\bigg)^{-(m+2)}.
\end{equation}
Therefore the metric \eqref{eq2} from Eqs. \eqref{eq31}-\eqref{eq32} is acquired the following form:
\begin{equation}\label{eq34}
ds^2=dt^2-\bigg(x^{m}\Bigg(\frac{k_{2}}{k_{1}} \sinh\big(k_{1}t+k_{3}\big)\Bigg)^{m}\bigg)^{2} dx^2-\Bigg(\frac{k_{2}}{k_{1}} \sinh\big(k_{1}t+k_{3}\big)\Bigg)^{2}(dy^2+\big(\sin y\big)^2dz^2).
\end{equation}
Further $V$, $a$ \& $H$ are defined as 
\begin{align}\label{eq35}
V=&x^{m}\bigg(\frac{k_{2}\sinh\big(k_{1}t+k_{3}\big)}{k_{1}}\bigg)^{m+2}\sin y,\\
\label{eq36}
a=&x^{\frac{m}{3}}\bigg(\frac{k_{2}\sinh\big(k_{1}t+k_{3}\big)}{k_{1}}\bigg)^{\frac{m+2}{3}}\big(\sin y\big)^{\frac{1}{3}},\\
\label{eq37}
\& \quad  H=&\frac{k_{1}(m+2) \coth\big(k_{1}t+k_{3}\big)}{3}.
\end{align}

From Eqs.\eqref{eq8} and \eqref{eq37}, the energy density of the model is given by
\begin{equation}\label{eq38}
\rho_{de}=\bigg(-1+\frac{2}{3}(m+2)\cosh^2\big(k_{1}t+k_{3}\big)\bigg)k_{1}^2(m+2) \alpha \csch^2\big(k_{1}t+k_{3}\big).
\end{equation}
The energy density of matter is given as
\begin{equation}\label{eq39}
\rho_{m}=\frac{\varPhi_{3}}{16 \pi k_{2}^4 \sinh^{4}\big(k_{1}t+k_{3}\big) \sin y },
\end{equation}
$$
\left.
\begin{aligned}
\text{where}\quad \varPhi_{3}=& \Bigg(k_{1}^{2}\Bigg( \bigg(\frac{k_{2} \sinh\big(k_{1}t+k_{3}\big)}{k_{1}}\bigg)^{-m} \bigg(k^{2}_{2}\bigg( \big(+4m+2\big)k_{1}^{2}-2 \big(m+2\big)^{2}k_{1}-\big(m+2\big)^{2}\varw\bigg)\\&\cosh^{2}\big(k_{1}t+k_{3}\big)
+2k^{2}_{1}\bigg) \phi_{0} x^{-m}- \bigg(16-\frac{32}{3}\big(m+2\big)\cosh^{2}\big(k_{1}t+k_{3}\big)\bigg) k_{2}^{4} \pi\\& \alpha \big(m+2\big) \sin y \sinh^{2}\big(k_{1}t+k_{3}\big)\Bigg)
\Bigg).
\end{aligned}\right\}
$$
The effective pressure is
\begin{equation}\label{eq40}
\overline{p}_{de}=\frac{\varPhi_{4}}{16 \pi k_{2}^4 \sinh^{6}\big(k_{1}t+k_{3}\big) \sin y},
\end{equation}
$$
\left.
\begin{aligned}
\text{where}\quad \varPhi_{4}=&\Bigg(\bigg(-2 k_{2}^{2}\bigg(1+(m+2)\cosh^{2}\big(k_{1}t+k_{2}\big)\bigg)(m+2) \sinh^{2}\big(k_{1}t+k_{2}\big)-\big(\cosh\big(k_{1}t+k_{2}\big)-1\big)\\& \bigg(\bigg(6k_{1}^{2}+(-4m-8)k_{1}+(m+2)^{2} \varw\bigg)k_{2}^{2} \cosh^{2}\big(k_{1}t+k_{2}\big)+\bigg(-4k_{2}^{2}+2\bigg)k_{1}^{2}\bigg(\cosh\big(k_{1}t+k_{2}\big)+1\bigg)\bigg)\\&\phi_{0}x^{-m}\bigg(\frac{k_{2}\sinh\big(k_{1}t+k_{3}\big)}{k_{1}}\bigg)^{-m}\bigg)\Bigg).
\end{aligned}\right\}
$$
The viscosity coefficient takes the form
\begin{equation}\label{eq41}
\xi=\frac{3 \cosh^2\big(k_{1}t+k_{3}\big) \xi_{0}+\xi_{1}k_{1} \sinh\big(k_{1}t+k_{3}\big)(m+2)\cosh\big(k_{1}t+k_{3}\big)-\xi_{2}k_{1}^{2}(m+2)-3\xi_{0}}{3 \sinh^2\big(k_{1}t+k_{3}\big)}.
\end{equation}
The proper pressure is 
\begin{equation}\label{eq42}
p_{de}=\frac{\varPhi_{5}}{3 \pi k_{2}^{4} \sin y \sinh^{6}\big(k_{1}t+k_{3}\big)},
\end{equation}
$$
\left.
\begin{aligned}
\text{where}\quad &\varPhi_{5}=\Bigg(k_{1} \sinh^{2}\big(k_{1}t+k_{3}\big)\Bigg(\frac{9 \phi_{0}}{8}\bigg(\bigg(\bigg(\frac{+2m}{3}+\frac{4}{3}\bigg)k_{1}-k_{1}^{2}-\frac{1}{6}(m+2)^{2} (\varw+2)\bigg)k_{2}^{2} \cosh^{2}\big(k_{1}t+k_{3}\big)\\&-\bigg(\frac{-2 k_{1}^{2}}{3}+\frac{m}{3}+\frac{2}{3}\bigg)k_{2}^{2}-\frac{k_{1}^{2}}{3}\bigg)  x^{-m} k_{1} \bigg(\frac{k_{2} \sinh\big(k_{1}t+k_{3}\big)}{k_{1}}\bigg)^{-m}-\bigg(-\xi_{0} k_{1} (m+2) \cosh^{3}\big(k_{1}t+k_{3}\big)\\&-3 \sinh\big(k_{1}t+k_{3}\big) \xi_{0} \cosh^{2}\big(k_{1}t+k_{3}\big)+\xi_{1} k_{1}(m+2)\cosh\big(k_{1}t+k_{3}\big)+\bigg(\xi_{2} k_{1}^{2}(m+2)+3\xi_{0}\bigg)\\&\sinh\big(k_{1}t+k_{3}\big)\bigg)\cosh\big(k_{1}t+k_{3}\big) \pi k_{2}^{4} (m+2) \sin y\Bigg)\Bigg).
\end{aligned}\right\}
$$
\begin{figure}[H]
	\centering
	\begin{minipage}{0.50\textwidth}
		\includegraphics[width=1\linewidth]{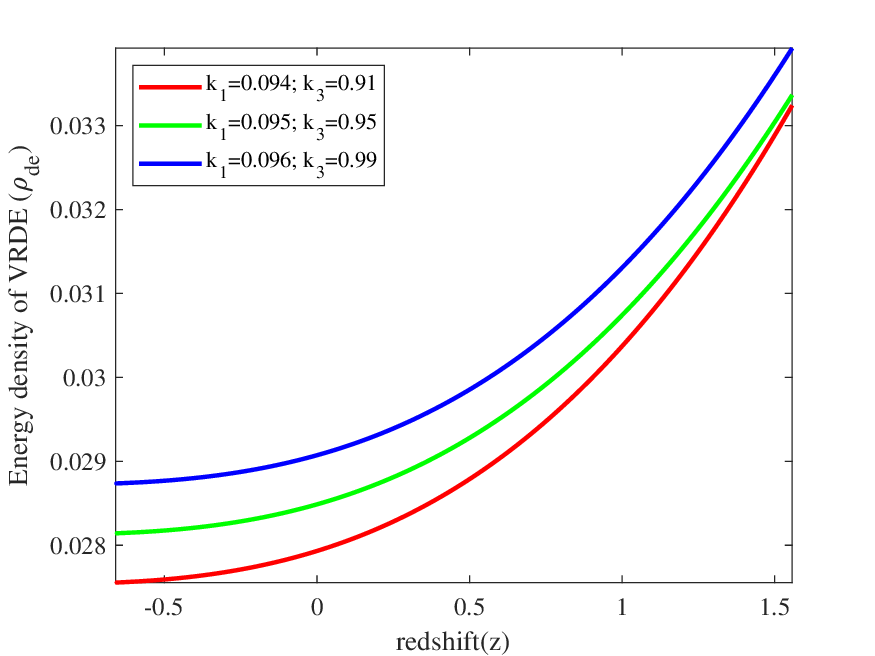}
		\caption{ Plot of energy density of VRDE\\ $(\rho_{de})$ versus redshift $(z)$.}
		\label{fig:rhode1}
	\end{minipage}
	\begin{minipage}{0.50\textwidth}
		\includegraphics[width=1\linewidth]{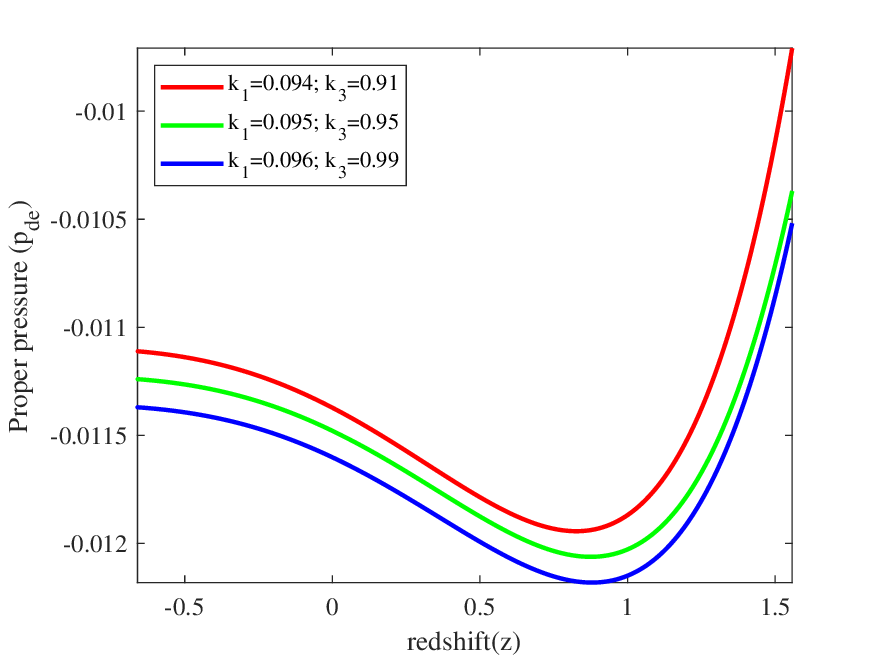}
		\caption{ Plot of proper pressure $(p_{de})$ versus redshift $(z)$.}
		\label{fig:pp1}
	\end{minipage}
\end{figure}
\subsection{Model--III}
If $\kappa=-1$, then from the Eq. \eqref{eq17} for $n=-3$, we obtain
\begin{equation}\label{eq43}
R=\Bigg(\frac{k_{5}}{k_{4}} \sinh\big(k_{4}t+k_{6}\big)\Bigg),
\end{equation}
and using Eq. \eqref{eq43} and Eq.\eqref{eq31}, we get
\begin{equation}\label{eq44}
Q=x^{m}\Bigg(\frac{k_{5}}{k_{4}} \sinh\big(k_{4}t+k_{6}\big)\Bigg)^{m}.
\end{equation}
The scalar field $\phi$ is given by 
\begin{equation}\label{eq45}
\phi=\phi_{0}x^{-m} \big(\sinh y\big)^{-1} \bigg(\frac{k_{5}\sinh\big(k_{4}t+k_{6}\big)}{k_{4}}\bigg)^{-(m+2)}.
\end{equation}
With the help of $R$ and $Q$, the metric \eqref{eq2} takes the form as
\begin{equation}\label{eq46}
ds^2=dt^2-\bigg(x^{m}\Bigg(\frac{k_{5}}{k_{4}} \sinh\big(k_{4}t+k_{6}\big)\Bigg)^{m}\bigg)^{2} dx^2-\Bigg(\frac{k_{5}}{k_{4}} \sinh\big(k_{4}t+k_{6}\big)\Bigg)^{2}(dy^2+\big(\sinh y\big)^2dz^2).
\end{equation}
The expressions of $V$, $a$ and $H$ are given by 
\begin{align}\label{eq47}
V=&x^{m}\bigg(\frac{k_{5}\sinh\big(k_{4}t+k_{6}\big)}{k_{4}}\bigg)^{m+2}\sinh y, \\
\label{eq48}
a=&x^{\frac{m}{3}}\bigg(\frac{k_{5}\sinh\big(k_{4}t+k_{6}\big)}{k_{4}}\bigg)^{\frac{m+2}{3}}\big(\sinh y\big)^{\frac{1}{3}},\\
\label{eq49}
\& \quad H=&\frac{k_{4}(m+2) \coth\big(k_{4}t+k_{6}\big)}{3}.
\end{align}
From Eqs.\eqref{eq8} and \eqref{eq49}, the energy density of the VRDE model is given by
\begin{equation}\label{eq50}
\rho_{de}=\bigg(-1+\frac{2}{3}(m+2)\cosh^2\big(k_{4}t+k_{6}\big)\bigg)k_{4}^2(m+2) \alpha \csch^2\big(k_{4}t+k_{6}\big).
\end{equation}
The energy density of matter is given by
\begin{equation}\label{eq51}
\rho_{m}=\frac{\varPhi_{6}}{48 \pi k_{5}^4 \sinh^{4}\big(k_{4}t+k_{6}\big) \sinh y  },
\end{equation}
$$
\left.
\begin{aligned}
\text{where}\quad \varPhi_{6}=& \Bigg(
3\phi_{0} x^{-m} k_{4}^{2}\Bigg(k_{5}^{2} \bigg(k_{4}^{2} (4m+2) k_{4}^{2}-2 (2+m)^{2} k_{4}-\varw (m+2)^{2}\bigg) \cosh^{2}\big(k_{4}t+k_{6}\big)\\&-2k_{4}^{2}\Bigg)\bigg(\frac{k_{5} \sinh\big(k_{4}t+k_{6}\big)}{k_{4}}\bigg)^{-m}+32 \pi \sinh y k_{5}^{4} \bigg(\frac{-3}{2}+(2+m)\cosh^{2}\big(k_{4}t+k_{6}\big)\bigg)\\& \sinh^{2}\big(k_{4}t+k_{6}\big) k_{5}^{2} \alpha (m+2)\Bigg),
\end{aligned}\right\}
$$
the effective pressure as
\begin{equation}\label{eq52}
\overline{p}_{de}=\frac{\varPhi_{7}}{16 \pi k_{5}^4 \sinh^{6}\big(k_{4}t+k_{6}\big) \sinh y},
\end{equation}
$$
\left.
\begin{aligned}
\text{where}\quad \varPhi_{7}=&\Bigg(\bigg(\frac{k_{5}\sinh\big(k_{4}t+k_{6}\big)}{k_{4}}\bigg)^{-m}\phi_{0}x^{-m}k_{4}^{2}\Bigg(-2\bigg(1+(2+m)\cosh^{2}\big(k_{4}t+k_{6}\big)\bigg)k_{5}^{2}(2+m)\sinh^{2}\big(k_{4}t+k_{6}\big)\\&-\sinh^{2}\big(k_{4}t+k_{6}\big)\bigg(6k_{4}^{2}+(-4m-8)k_{4}+\varw(2+m)^{2}\bigg)k_{5}^{2}\cosh^{2}\big(k_{4}t+k_{6}\big)+\bigg(+4k_{5}^{2}+2
\bigg)k_{4}^{2}\Bigg)\Bigg),
\end{aligned}\right\}
$$
the viscosity coefficient as
\begin{equation}\label{eq53}
\xi=\frac{3 \cosh^2\big(k_{4}t+k_{6}\big) \xi_{0}+\xi_{1}k_{4} \sinh\big(k_{4}t+k_{6}\big)(m+2)\cosh\big(k_{4}t+k_{6}\big)-\xi_{2}k_{4}^{2}(m+2)-3\xi_{0}}{3 \sinh^2\big(k_{4}t+k_{6}\big)}, 
\end{equation}
and the proper pressure is given by
\begin{equation}\label{eq54}
p_{de}=\frac{3\varPhi_{8}}{8 \pi k_{5}^{4} \sinh y \sinh^{6}\big(k_{4}t+k_{6}\big)},
\end{equation}
$$
\left.
\begin{aligned}
\text{where}\quad \varPhi_{8}=&\Bigg(k_{4} \sinh^{2}\big(k_{4}t+k_{6}\big)\Bigg(\bigg(k_{5}^{2}\bigg( \bigg(\frac{2m}{3}+\frac{4}{3}\bigg)k_{4}-k_{4}^{2}-\frac{1}{6}(m+2)^{2} (\varw+2)\bigg)\cosh^{2}\big(k_{4}t+k_{6}\big)\\&-\bigg(\frac{-2 k_{4}^{2}}{3}+\frac{m}{3}+\frac{2}{3}\bigg)k_{5}^{2}+\frac{k_{4}^{2}}{3}\bigg) x^{-m} k_{4} \phi_{0} \bigg(\frac{k_{5} \sinh\big(k_{4}t+k_{6}\big)}{k_{4}}\bigg)^{-m}-\frac{8}{9}\cosh\big(k_{4}t+k_{6}\big) \pi\\& k_{5}^{4} (m+2) \sinh y  \bigg(
-\xi_{1} k_{4} (m+2) \cosh^{3}\big(k_{4}t+k_{6}\big)-3 \sinh\big(k_{4}t+k_{6}\big) \xi_{0} \cosh^{2}\big(k_{4}t+k_{6}\big)\\&+\xi_{1} k_{4}(m+2)\cosh\big(k_{4}t+k_{6}\big)+\bigg(\xi_{2} k_{4}^{2}(m+2)+3\xi_{0}\bigg)\sinh\big(k_{4}t+k_{6}\big)\bigg)\bigg)\Bigg).
\end{aligned}\right\}
$$
\begin{figure}[H]
	\centering
	\begin{minipage}{0.50\textwidth}
		\includegraphics[width=1\linewidth]{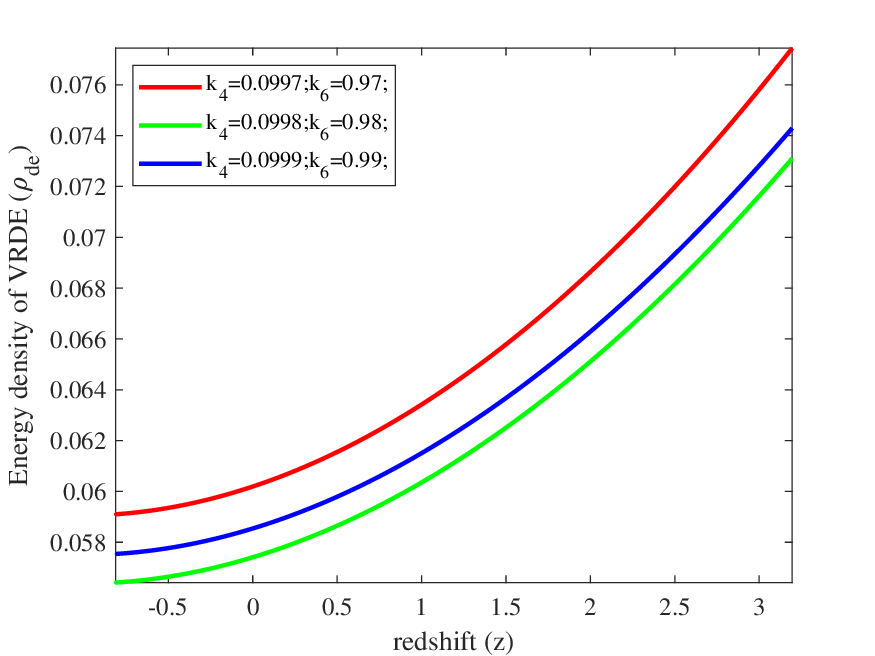}
		\caption{ Plot of energy density of VRDE \\$(\rho_{de})$ versus redshift $(z)$.}
		\label{fig:rhodem1}
	\end{minipage}
	\begin{minipage}{0.50\textwidth}
		\includegraphics[width=1\linewidth]{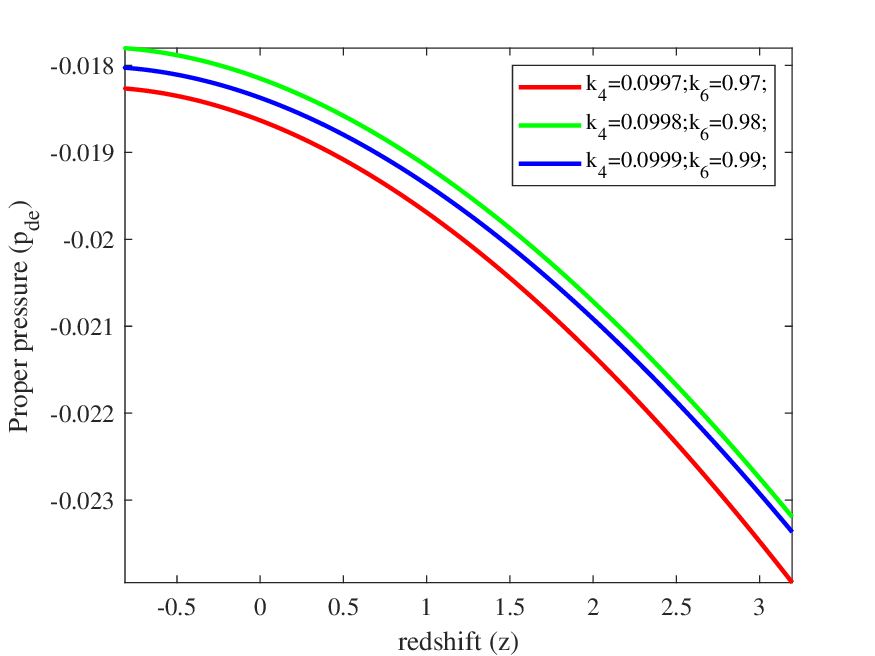}
		\caption{ Plot of proper pressure $(p_{de})$ versus redshift $(z)$.}
		\label{fig:ppm1}
	\end{minipage}
\end{figure}
The ordinary matter species namely baryons and radiation; and the DE are two incomparable quantities as DE possess the negative pressure that accelerates the expansion of the Universe by restraining the gravitational force. We have plotted the energy density of VRDE ($\rho_{de}$) versus redshift $(z)$ for three models with respect to the values of $c_{1}$, $c_{2}$, $k_{1}$, $k_{3}$, $k_{4}$, $k_{6}$ in figures \eqref{fig:rhode0}, \eqref{fig:rhode1} \& \eqref{fig:rhodem1} respectively and we can observe that the curves of $\rho_{de}$ varying in the positive region throughout the evolution of the Universe against redshift $(z)$, which indicates the Universe expansion in an accelerated way. Figures \eqref{fig:pp0}, \eqref{fig:pp1} \& \eqref{fig:ppm1} respectively represent the trajectories of proper pressure ($p_{de}$) against redshift ($z$) for all three models with different values of $c_{1}$, $c_{2}$, $k_{1}$, $k_{3}$, $k_{4}$, and $k_{6}$. We observe that the path of the proper pressure for three different values of $c_{1}$, $c_{2}$, $k_{1}$, $k_{3}$, $k_{4}$, \& $k_{6}$ traverse in negative region. Here, the occurrence of the Universe's accelerated expansion can be implied from this negative pressure in the BDT.\\ Also, Eqs. \eqref{eq21}, \eqref{eq34} and \eqref{eq46} represent Ruban's VRDE models in BDT of gravity for $\kappa=0,$\,$1$\, and $-1$ ($i.e.,$ models--I, II \& III) respectively along with the above discussed and following properties. 
\begin{itemize}
	\item The expansion scalar $(\theta)$ and shear scalar $(\sigma^2)$ for the models-I, II \& III respectively given as
	\begin{align}
	\theta=&\bigg(\frac{\dot{Q}}{Q}+2\frac{\dot{R}}{R}\bigg), \quad \sigma^2=\frac{1}{2}\bigg(\sum_{\mu=1}^{3}H_\mu^2-\frac{\theta^2}{3}\bigg), \\
	\theta=&\frac{c_1(2+m)}{\gamma\big(tc_1+c_2\big)},\quad\sigma^2=\frac{c_{1}^{2}(m-1)^{2}}{3\big(\gamma(c_{1}t+c_{2})\big)^{2}},\\
	\theta=&(m+2)k_{1}\coth\big(k_{1}t+k_{3}\big),\quad \sigma^2=\frac{k_{1}^{2} \coth^{2}\big(k_{1}t+k_{3}\big)\big(m-1\big)^{2}}{3}, \\
	\& \quad \theta=&(m+2)k_{4}\coth\big(k_{4}t+k_{6}\big),\quad \sigma^2=\frac{k_{4}^{2} \coth^{2}\big(k_{4}t+k_{6}\big)\big(m-1\big)^{2}}{3}.
	\end{align}
	\item The anisotropic parameter of the models--I, II \& III is same and is given by 
	\begin{equation}
	\mathcal{A}_h=\frac{1}{3}\sum_{\mu=1}^{3}\left(\frac{H_\mu-H}{H}\right)^2=\frac{2(m-1)^2}{(2+m)^2}.
	\end{equation}	
	It is noted that for all three models, the physical quantities $\theta$, $\sigma^2$ and $H$ tend to infinity at $t=\frac{-c_2}{c_1},$ $t=\frac{-k_3}{k_1}$ \& $t=\frac{-k_6}{k_4}$ respectively, and the scale factor of the models, as well as the spatial volume vanishes. However, these parameters become constant as $t \rightarrow \infty$ and the scale factor, as well as volume tend to infinity as time increases. Therefore, this summarizes that the models show expansion with zero volume in the beginning period and further expand to infinitely large $V$ w.r.t. cosmic time $t$(Gyr). As $\mathcal{A}_h\neq0$ for all three models, thus the models are anisotropic throughout the evolution of the Universe.
\end{itemize}
\section{Analysis of geometrical parameters}\label{c8agp}
This segment of the work looks through the expanding behavior of the cosmos by studying well known astronomical parameters like the deceleration parameter$(q)$, jerk parameter $(j)$, $r-s$ plane, $q-r$ plane, EoS parameter($\omega_{de}$),  $\omega_{de}-\omega_{de}'$ plane, the stability of the model $(v_s^2)$, and om-diagnostic for the constructed VRDE models which are in Eqs. \eqref{eq21}, \eqref{eq34} and \eqref{eq46}.  
\begin{itemize}
	\item {\bf Deceleration parameter:} The deceleration parameter (DP) is defined as
	\begin{equation}
	q=-\frac{a\ddot{a}}{\dot{a}^2} ,
	\end{equation}
	that depends upon the scale factor and its derivatives, which can be considered to explain the transition phase of the cosmos and it basically computes the expansion rate of the cosmos. Whenever the DP shows the positive curve, it indicates the decelerated expansion of the Universe. Whereas, the negative curve implies that there is an accelerated expansion of the cosmos, and at $q=0$ there exists marginal inflation. For the constructed models--I, II \& III the DP takes the values as:	
	\begin{align}\label{eq61}
	q=&\frac{3\gamma}{2+m}-1, \\
	\label{eq62}
	q=&-1+3(m+2)\sech^2\big(k_{1}t+k_{3}\big),\\
	\label{eq63}
	\& \quad  q=&-1+3(m+2)\sech^2\big(k_{4}t+k_{6}\big).
	\end{align}
	The behavior of the deceleration parameter($q$) is addressed in figures \eqref{fig:q1} \& \eqref{fig:qm1} against redshift ($z$) for models--II \& III with different values of $k_{1}$, $k_{3}$, $k_{4}$, \& $k_{6}$. The path of the curves for the deceleration parameter travels from the early decelerated phase to the present accelerated phase of the Universe agreeing with the recent astrophysical calculations, whereas the deceleration parameter for model--I is independent of time and Santhi and Naidu \cite{newtcn3,newtcn4,tcn2} have obtained constant deceleration parameter in the literature.
	\begin{figure}[H]
		\centering
		\begin{minipage}{0.49\textwidth}
			\includegraphics[width=1\linewidth]{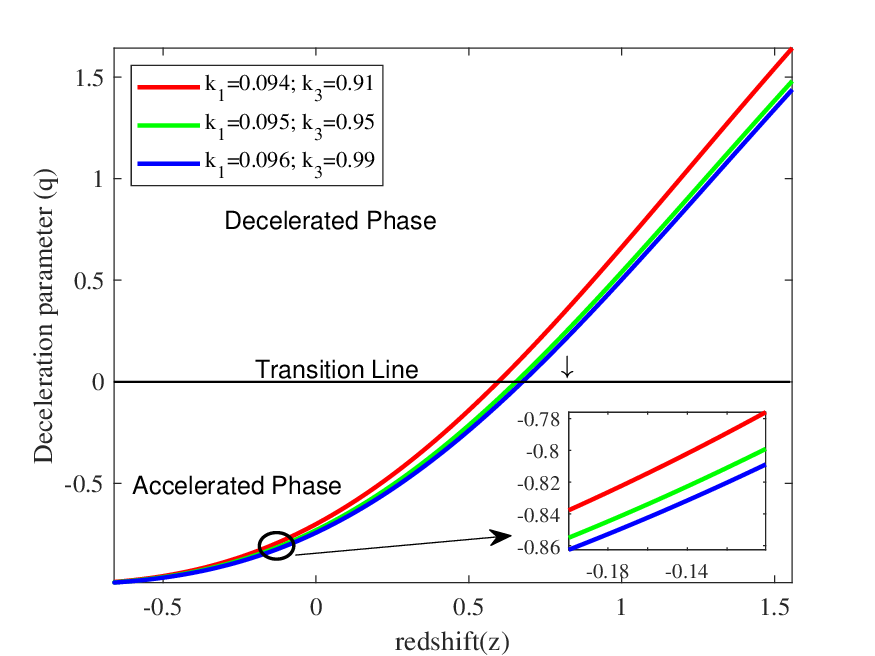}
			\caption{Plot of DP $(q)$ versus\\ redshift $(z)$ for Model--II.}
			\label{fig:q1}
		\end{minipage}
		\begin{minipage}{0.49\textwidth}
			\includegraphics[width=1\linewidth]{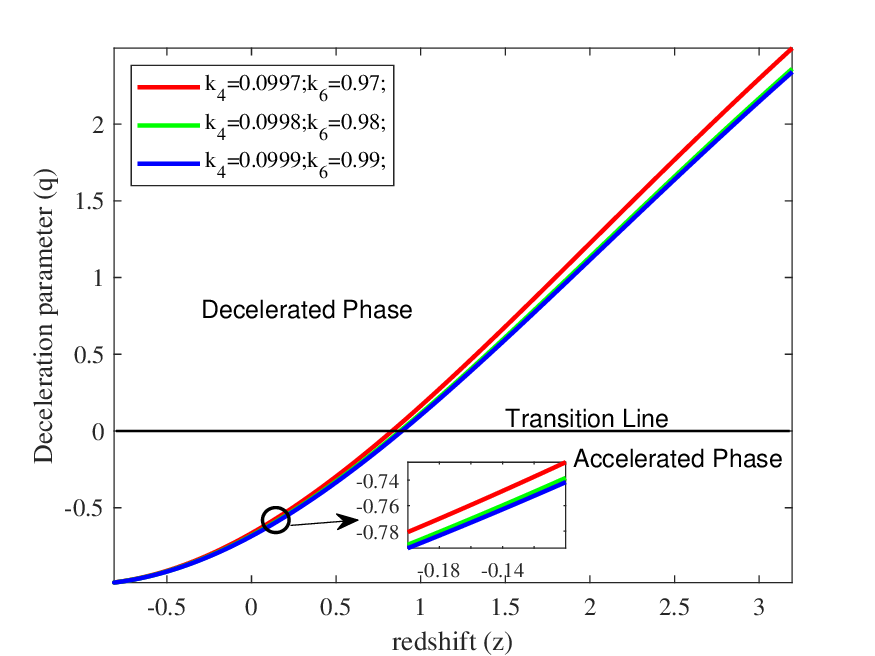}
			\caption{Plot of DP $(q)$ versus\\ redshift $(z)$ for Model--III.}
			\label{fig:qm1}
		\end{minipage}
	\end{figure}
	\item {\bf Jerk parameter:} 
	A dimensionless cosmic jerk parameter is obtained by the third derivative of the average scale factor w.r.t. cosmic time `t', which is given by
	\begin{equation}\label{tw50}
	j=\frac{\dddot{a}}{aH^{3}}=q\left(1+2q\right)-\frac{\dot{q}}{H}\,\,.
	\end{equation}
	Cosmic jerk can be accounted for the transition of the Universe from decelerating to the accelerating phase. For various models of the cosmos, there is a variation in the transition of the cosmos, whenever the jerk parameter lies in the positive region and the negative values of DP (Visser \cite{Vis}). The investigations of Rapetti et al. \cite{Rap} have shown that for a flat $\Lambda$CDM model, the value of jerk becomes unity. The jerk parameter for models--I, II \& III is given by 
	\begin{align}\label{eq65}
	j=&\frac{(-3\gamma+2+m)(2+m-6\gamma)}{(2+m)^2},
	\\
	\label{eq66}
	j=&1-\frac{9m}{(m+2)^{2}}\sech^2\big(k_{1}t+k_{3}\big),\\
	\label{eq67}
	\& \quad j=&1-\frac{9m}{(m+2)^{2}}\sech^2\big(k_{4}t+k_{6}\big).
	\end{align}
	\begin{figure}[H]
		\centering
		\begin{minipage}{0.50\textwidth}
			\includegraphics[width=1\linewidth]{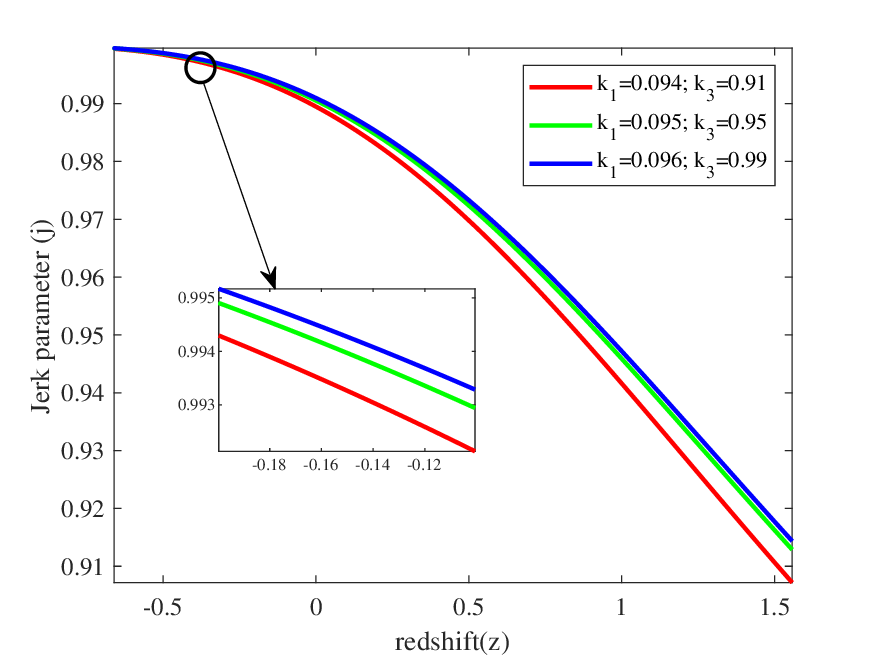}
			\caption{Plot of jerk parameter $(j)$ \\versus redshift $(z)$ for Model--II.}
			\label{fig:jerk1}
		\end{minipage}
		\begin{minipage}{0.50\textwidth}
			\includegraphics[width=1\linewidth]{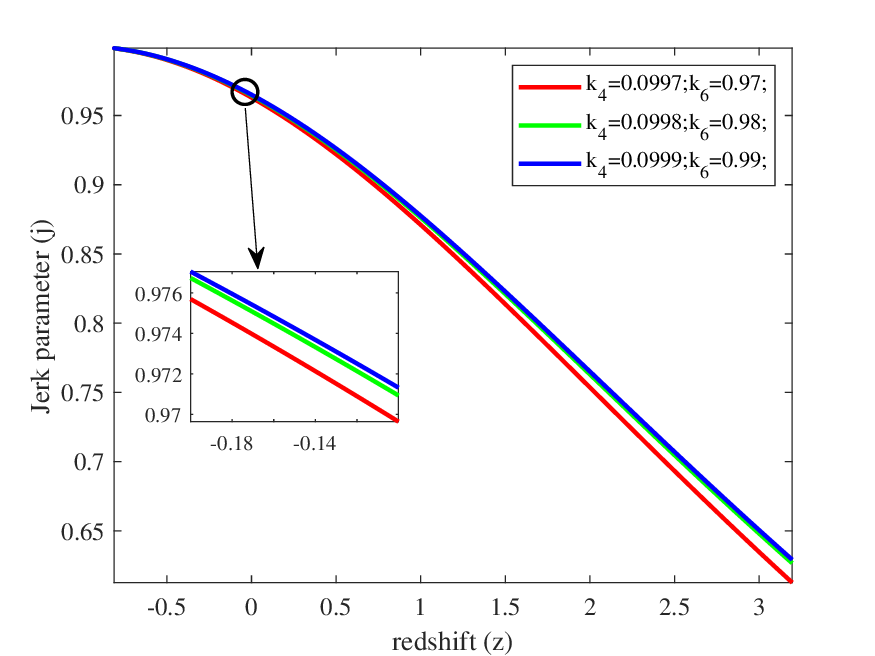}
			\caption{Plot of jerk parameter $(j)$ versus redshift $(z)$ for Model--III.}
			\label{fig:jerkm1}
		\end{minipage}
	\end{figure}
	The plots for jerk parameter ($j$) against redshift are represented graphically in figures \eqref{fig:jerk1} \& \eqref{fig:jerkm1} for models-II \& III with different values of $k_{1}$, $k_{3}$, $k_{4}$, \& $k_{6}$ respectively. It can be analyzed from the plots that the jerk parameter for both the models differ in the positive regions and approaches one in near future as $z \rightarrow 0$, whereas jerk parameter for model-I is independent of time. 
	\item {\bf Statefinder parameters:} A recently developed geometrical diagnostic is this statefinder pair $(r,s)$ given by Sahni et al. \cite{sah} and Alam et al. \cite{ala} who have proposed for the purpose of distinction among the various DE candidates. This is a sensitive and geometrical diagnostic pair that is essential to discriminate and study the diverse DE models and hence help us interpret cosmic acceleration. This geometrical pair is represented as ($r,s$), and are formulated as
	\begin{equation}
	r=\frac{\dddot{a}}{aH^3} \quad\text{and}\quad s=\frac{r-1}{3(q-\frac{1}{2})}.
	\end{equation}
	The $r-s$ plane helps to analyze several cases of the model incorporating different parameters and spacial curvature components. For different models of DE, we have different evolutionary trajectories in $r-s$ plane as for $(r,s)=(1,0)$ we obtain $\Lambda$CDM model and $(r,s)=(1,1)$ relates to SCDM model. The statefinder parameters ($r,s$) for our VRDE models--I, II \& III are obtained as
	\begin{align}\label{eq69}
	r=&\frac{(-3\gamma+2+m)(2+m-6\gamma)}{(2+m)^2};\quad s=\frac{2\gamma}{2+m}, \\ \label{eq70}
	r=&1-\frac{9m}{(m+2)^{2}}\sech^2\big(k_{1}t+k_{3}\big);\quad s=\frac{2m}{(m+2)^{2}\bigg(\cosh^{2}\big(k_{1}t+k_{3}\big)-2m-4\bigg)},\\\label{eq71}
	\& \quad r=&1-\frac{9m}{(m+2)^{2}}\sech^2\big(k_{4}t+k_{6}\big);\quad s=\frac{2m}{(m+2)^{2}\bigg(\cosh^{2}\big(k_{4}t+k_{6}\big)-2m-4\bigg)}.
	\end{align}
The trajectories of $r-s$ and $q-r$ planes are plotted to evaluate different DE models. The quantities $q$, $r$ and $s$ are model independent because they depend only on $`a'$ and its derivatives of higher order. Hence this approach doesn't need understanding of gravity theory. Here, the steady state (SS) model is represented by $(q,r)=(-1,1)$ and the SCDM model is represented by $(q,r) = (0.5,1)$.
	\begin{figure}[H]
		\centering
		\begin{minipage}{0.50\textwidth}
			\includegraphics[width=1\linewidth]{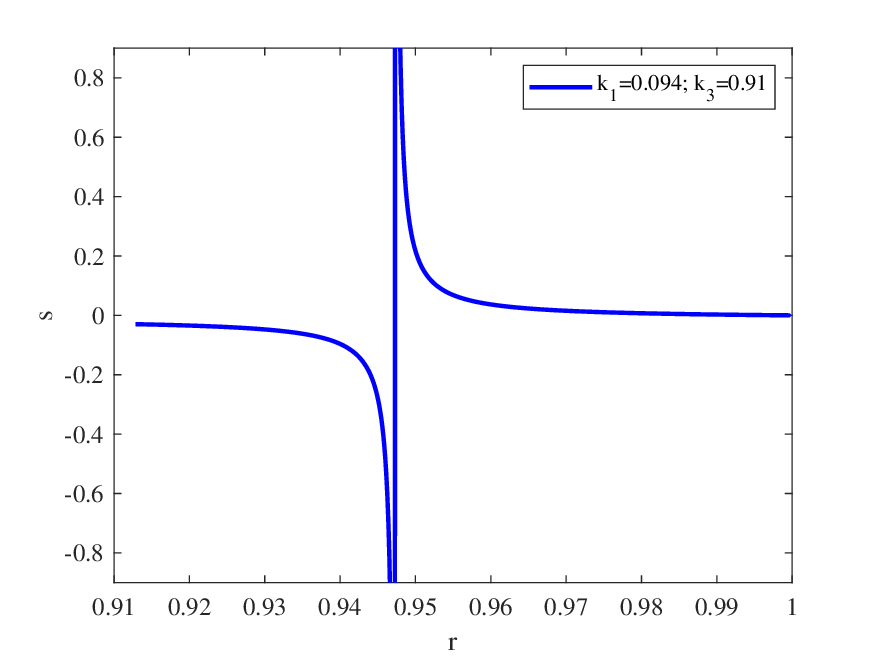}
			\caption{ Plot of $r-s$ plane : Model--II.}
			\label{fig:rs1}
		\end{minipage}
		\begin{minipage}{0.50\textwidth}
			\includegraphics[width=1\linewidth]{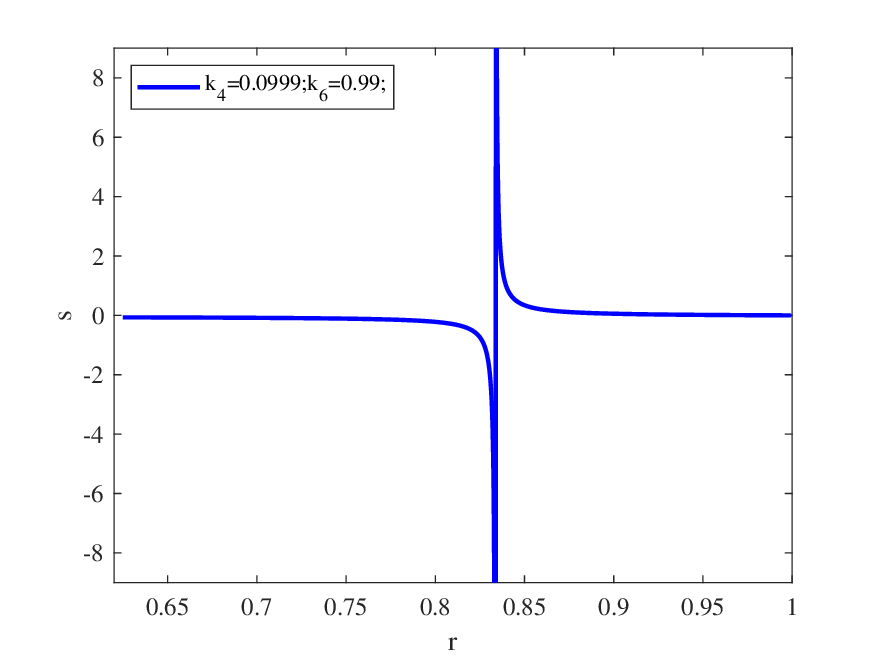}
			\caption{ Plot of $r-s$ plane : Model--III.}
			\label{fig:rsm1}
		\end{minipage}
	\end{figure}
	\begin{figure}[H]
		\centering
		\begin{minipage}{0.50\textwidth}
			\includegraphics[width=1\linewidth]{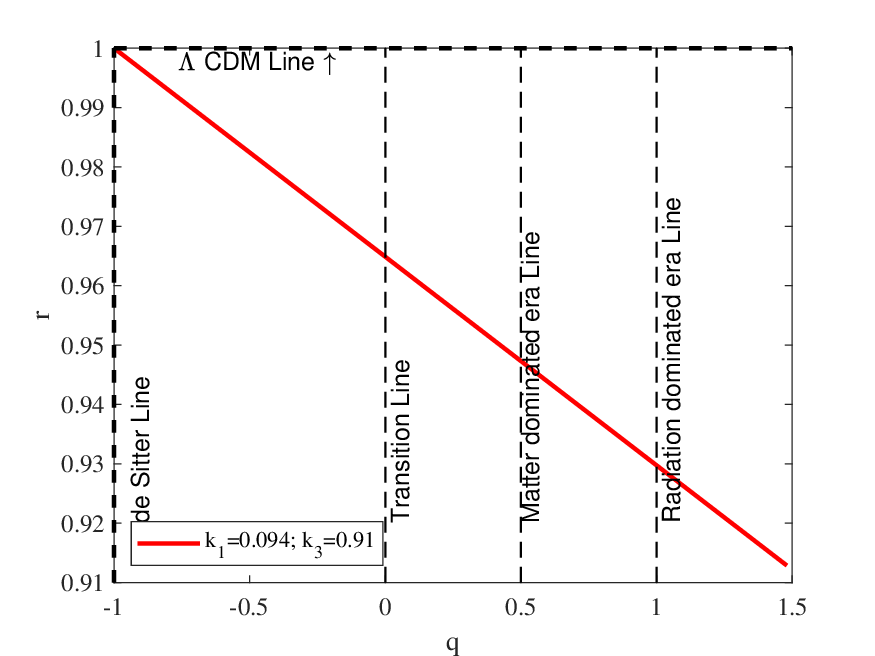}
			\caption{ Plot of $q-r$ plane : Model--II.}
			\label{fig:qr}
		\end{minipage}
		\begin{minipage}{0.50\textwidth}
			\includegraphics[width=1\linewidth]{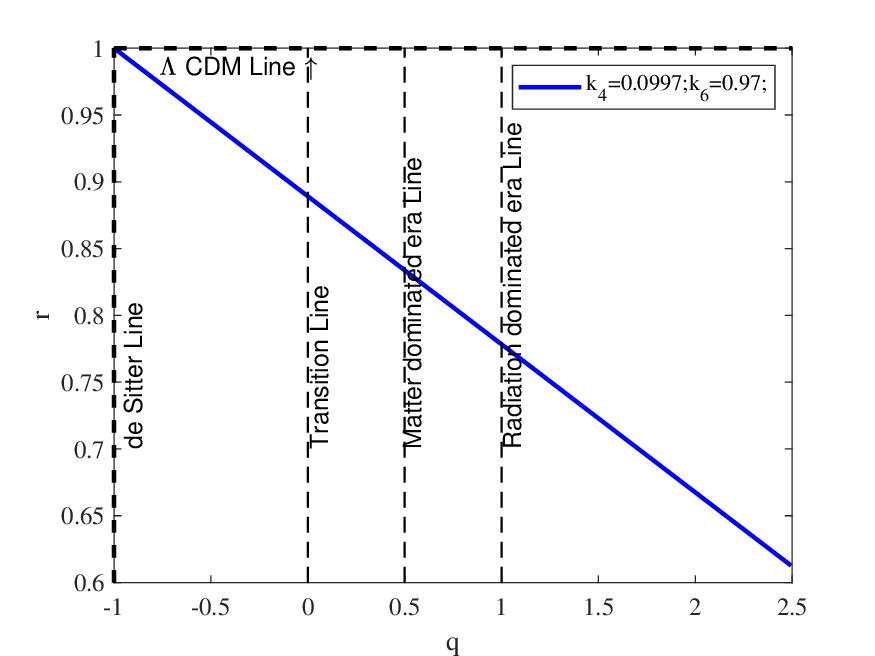}
			\caption{ Plot of $q-r$ plane : Model--III.}
			\label{fig:qrm1}
		\end{minipage}
	\end{figure}
	To understand the phase transition of the Universe, we have constructed the plots of $r-s$ \& $q-r$ planes. Figures \eqref{fig:rs1} and \eqref{fig:rsm1} represent the statefinder pair ($r,s$) for models--II \& III respectively. It could be interpreted from these figures that the models start their evolution from the quintessence and phantom region and reaches the $\Lambda$CDM model (for $r=1,s=0$). Figures \eqref{fig:qr} and \eqref{fig:qrm1} for models-II \& III respectively are the plots of $q-r$ plane. It can be seen from the figures that the path of $q-r$ plane shows a signature change from negative region to positive region i.e., the trajectories are traveling from radiation dominated era, passing through the matter dominated region and transition line then reaching the de-sitter phase of the cosmos.			
	\item {\bf EoS parameter:} To classify the phases of the inflating cosmos, viz. transition from decelerated to accelerated phases containing DE and radiation dominated eras, the EoS parameter ($\omega_{de}$) can be broadly used, whose expression is given as $\omega_{de}=\frac{p_{de}}{\rho_{de}}$. It categorizes various epochs as follows:\\
	\underline{Decelerated phase:}
	\begin{itemize}
		\item {stiff fluid ($\omega_{de}=1$)},
		\item {the radiation dominated phase ($0<\omega_{de}<\frac{1}{3}$)} and
		\item {dust fluid phase or cold dark matter ($\omega_{de}=0$)}.	
	\end{itemize}
	\underline{Accelerated phase:}
	\begin{itemize}
		\item {the quintessence phase ($-1<\omega_{de}<\frac{-1}{3}$)},
		\item {cosmological constant/vacuum phase ($\omega_{de}=-1$)} and
		\item {quintom era and phantom era ($\omega_{de}<-1$)}. 	
	\end{itemize}
	The EoS parameter for the obtained models are given by 
	\begin{align}\label{eq72}
	\omega_{de}=&\frac{\varPhi_{9}}{96\big(tc_1+c_2\big)\pi c_1(2+m)\alpha\big(m-\frac{3}{2}\gamma+2\big)},\\
	\label{eq73}
	\omega_{de}=&\frac{\varPhi_{10}}{k_{2}^{4} \sinh^{4}\big(k_{1}t+k_{3}\big)\pi \alpha k_{1} (m+2)\bigg(\frac{-3}{2}+(m+2) \cosh^{2}\big(k_{1}t+k_{3}\big)\bigg)\sin y}, \\
	\label{eq74}
	\& \quad \omega_{de}=&\frac{\varPhi_{12}}{k_{5}^{4} \sinh^{4}\big(k_{4}t+k_{6}\big) \pi \alpha k_{4}^{4} (m+2) \sinh y \bigg(\frac{-3}{2}+(m+2) \cosh^{2}\big(k_{4}t+k_{6}\big)\bigg)}, 
	\end{align}
	$$
	\left.
	\begin{aligned}
	\text{where}\quad \varPhi_{9}=&\bigg(-\big(tc_1+c_2\big)c_1\big(n^2(\varw+2)m^2+4\big((\varw+2)n-\frac{3}{2}\gamma+3\big)nm+(-12\gamma+24)n-36\gamma+54\big)\\&\phi_0\big(x^{\frac{m}{3}} y^{\frac{1}{3}} \big(\gamma\big(tc_1+c_2\big)\big)^{\frac{1}{3}\frac{(2+m)}{\gamma}}\big)^n+(4\varw+8)n^2+48\pi\big(\big((t\xi_{1}-\xi_{2})m+3\xi_{0}t^2\gamma+2\xi_{1}t-2\xi_{2}\big)c_1^2\\&+c_2(6\gamma t\xi_{0}+m\xi_{1}+2\xi_{1})c_1+3\xi_{0}c_2^2\gamma\big)(2+m)\bigg),
	\end{aligned}\right\}
	$$
	$$
	\left.
	\begin{aligned}
	\varPhi_{10}=&\Bigg(\frac{1}{2}\Bigg(\sinh^{2}\big(k_{1}t+k_{3}\big)\Bigg(\frac{-9 \phi_{0}}{8}\bigg(\bigg(k_{1}^{2}+\bigg(\frac{-2m}{3}-\frac{4}{3}\bigg)k_{1}+\frac{1}{6}(m+2)^{2}(\varw+2)\bigg)k_{2}^{2} \cosh^{2}\big(k_{1}t+k_{3}\big)+\bigg(\frac{-2 k_{1}^{2}}{3}\\&+\frac{m}{3}+\frac{2}{3}\bigg)k_{2}^{2}+\frac{k_{1}^{2}}{3}\bigg)x^{-m} k_{1} \bigg(\frac{k_{2} \sinh\big(k_{1}t+k_{3}\big)}{k_{1}}\bigg)^{-m}+ k_{2}^{4} \pi  (m+2)\sin y\cosh\big(k_{1}t+k_{3}\big)\Bigg(\xi_{1}k_{1}(m+2)\\& \cosh^{3}\big(k_{1}t+k_{3}\big)+3 \sinh\big(k_{1}t+k_{3}\big) \xi_{0} \cosh^{2}\big(k_{1}t+k_{3}\big)-\xi_{1} k_{1}(m+2) \cosh\big(k_{1}t+k_{3}\big)\\&-\bigg(\xi_{2} k_{1}^{2}(m+2)+3\xi_{0}\bigg)\sinh\big(k_{1}t+k_{3}\big)\Bigg) \Bigg)\Bigg)\Bigg), 
	\end{aligned}\right\}
	$$
	$$
	\left.
	\begin{aligned}
	\& \quad \varPhi_{11}=&\Bigg(\frac{-1}{2}\Bigg(\sinh^{2}\big(k_{4}t+k_{6}\big)\Bigg(\frac{9 \phi_{0}x^{-m} k_{4} }{8}\bigg(\frac{k_{5} \sinh\big(k_{4}t+k_{6}\big)}{k_{4}}\bigg)^{-m}\bigg(k_{5}^{2}\bigg(k_{4}^{2}+\bigg(\frac{-2m}{3}-\frac{4}{3}\bigg)k_{4}\\&+\frac{1}{6}(m+2)^{2} (\varw+2)\bigg) \cosh^{2}\big(k_{4}t+k_{6}\big)+\bigg(\frac{-2 k_{4}^{2}}{3}+\frac{m}{3}+\frac{2}{3}\bigg)k_{5}^{2}-\frac{k_{4}^{2}}{3}+ k_{5}^{4} \pi  (m+2)\\&\sinh y\cosh\big(k_{4}t+k_{6}\big)\bigg)\bigg(
	-\xi_{1} k_{4} (m+2) \cosh^{3}\big(k_{4}t+k_{6}\big)-3 \sinh\big(k_{4}t+k_{6}\big)\\& \xi_{0} \cosh^{2}\big(k_{4}t+k_{6}\big)+\xi_{1} k_{4}(m+2)\cosh\big(k_{4}t+k_{6}\big)+\bigg(\xi_{2} k_{4}^{2}(m+2)+3\xi_{0}\bigg)\sinh\big(k_{4}t+k_{6}\big)\bigg)\Bigg) \Bigg)\Bigg). 
	\end{aligned}\right\}
	$$
	\begin{figure}[H]
		\centering
		\begin{minipage}{0.32\textwidth}
			\includegraphics[width=1\linewidth]{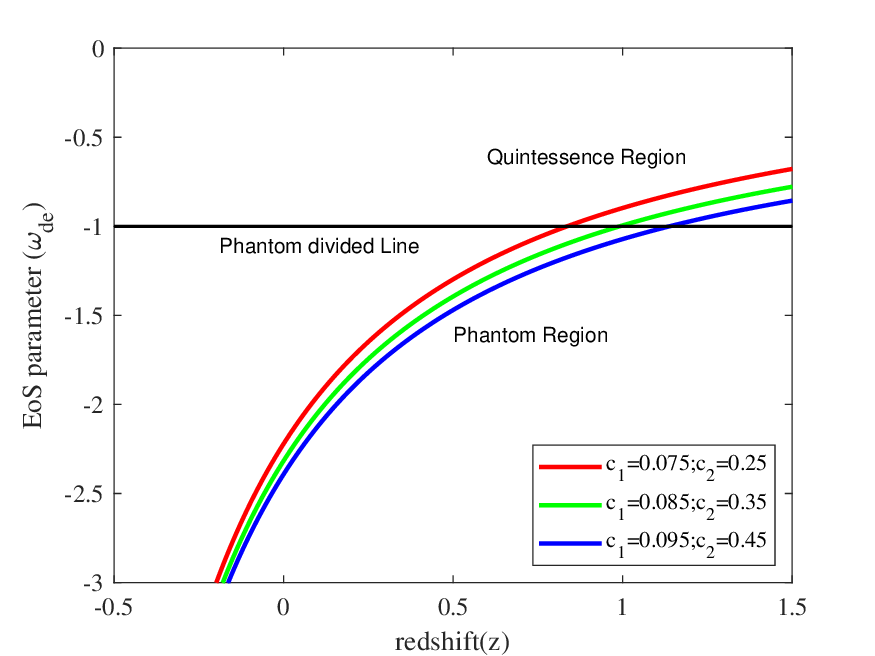}
			\caption*{Model--I}
		\end{minipage}
		\begin{minipage}{0.32\textwidth}
			\includegraphics[width=1\linewidth]{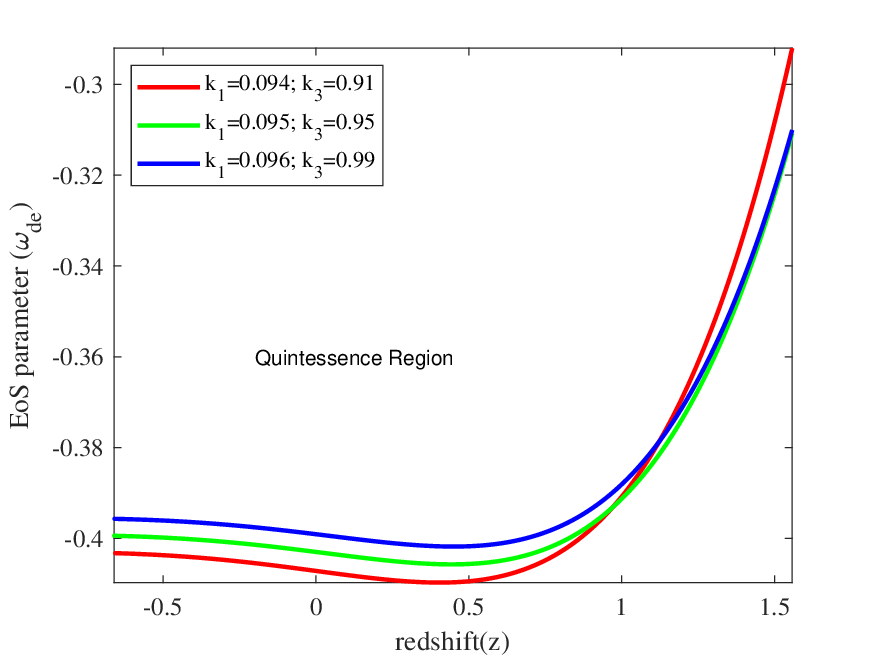}
			\caption*{Model--II}
		\end{minipage}
		\begin{minipage}{0.32\textwidth}
			\includegraphics[width=1\linewidth]{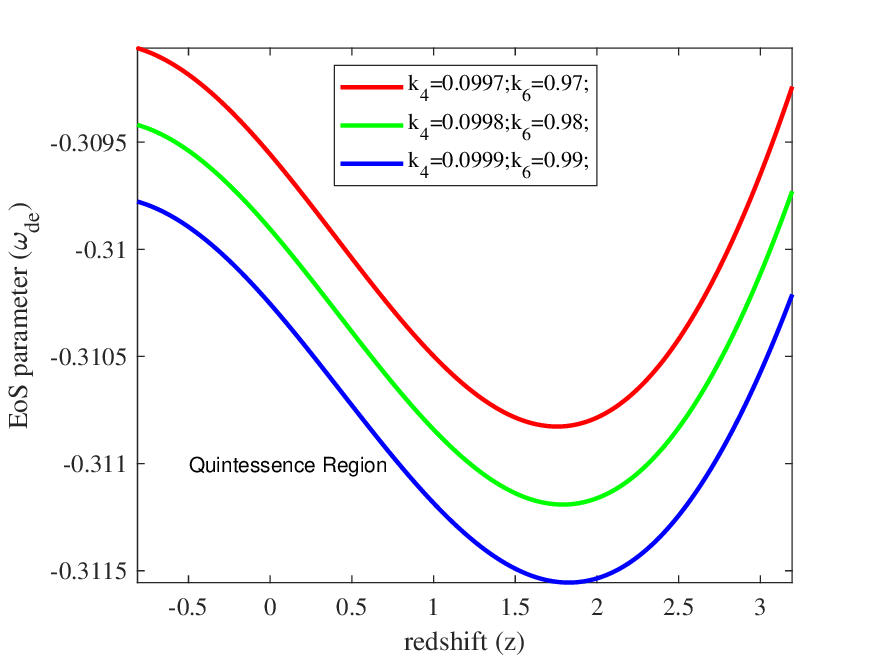}
			\caption*{Model--III}
		\end{minipage}
		\caption{ Plot of EoS parameter $(\omega_{de})$ versus redshift $(z)$.}
		\label{fig:eos01m1}
	\end{figure}
	Figure \eqref{fig:eos01m1} represents the plots of EoS parameter ($\omega_{de}$) against redshift $(z)$ for three models respectively with various values of $c_{1}$, $c_{2}$, $k_{1}$, $k_{3}$, $k_{4}$, \& $k_{6}$. Here, we observe that from the figure \eqref{fig:eos01m1} for model--I, the trajectories of $\omega_{de}$ travel from quintessence to phantom region, by crossing the phantom divided line, showing the quintom like behavior of the Universe and whereas, for model--II \& III, it is observed that the path of $\omega_{de}$ completely varies in the quintessence region representing the quintessence nature. Planck collaboration data $2018$ given by \cite{agh} for the EoS parameter, are consistent with the results of our models, where the limits of EoS parameter are given as follows:
	\begin{equation}\nonumber
	\omega_{de}=\left\{
	\begin{aligned}
	&-1.56_{-0.48}^{+0.60} (Planck + TT + lo\varw E)\\&  -1.58_{-0.41}^{+0.52} (Planck + TT, EE + lo\varw E)\\&  -1.57_{-0.40}^{+0.50} (Planck + TT, TE, EE + lo\varw E+lensing)\\&
	-1.04_{-0.10}^{+0.10} (Planck + TT, TE, EE + lo\varw E+lensing+BAO).
	\end{aligned}\right.
	\end{equation}
	\item {\bf {$\omega_{de}-\omega_{de}'$ plane:}} Cadwell and Linder \cite{cl} have suggested the $\omega_{de}-\omega_{de}'$ plane (where $'$ signifies differentiation w.r.t. $\ln$ a) to interpret the accelerated expansion regions of the cosmos and to analyze the quintessence scalar field for the first time. For various values of $\omega_{de}$ and $\omega_{de}'$, the plane describes two distinct areas. The plane is described as the thawing zone for $\omega_{de}'>0$ when $\omega_{de}<0$ and the freezing region for $\omega_{de}'<0$ when $\omega_{de}<0$. Also, the $\omega_{de}'$ expression for models--I, II \& III is given as
	\begin{align}\label{eq75}
	\omega'_{de}=&\frac{\varPhi_{12}}{96\big(tc_1+c_2\big)\pi c_1 (2+m) \alpha \big(m-\frac{3}{2}\gamma+2\big)},\\
	\label{eq76}
	\omega_{de}'=&\frac{\varPhi_{13}}{\Bigg(k_{2}^{4} \sinh^{4}\big(k_{1}t+k_{3}\big)\pi \alpha k_{1}^{2} (m+2)^{2} \sin y\, \cosh\big(k_{1}t+k_{3}\big) \bigg(\frac{-3}{2}+(m+2) \cosh^{2}\big(k_{1}t+k_{3}\big)\bigg)^{2} \Bigg)}, \\
	\label{eq77}
	\& \quad \omega_{de}'=&\frac{\varPhi_{14}}{k_{4}^{2}\bigg(\frac{-3}{2}+(m+2) \cosh^{2}\big(k_{4}t+k_{6}\big)\bigg)^{2}k_{5}^{4} \sinh^{4}\big(k_{4}t+k_{6}\big)\pi \alpha (m+2)^{2} \sinh y\, \cosh\big(k_{4}t+k_{6}\big)},
	\end{align}
		$$
	\left.
	\begin{aligned}
	\text{where}\quad \varPhi_{12}=&\bigg(-\big(tc_1+c_2\big) \phi_0 n c_1 \big((2+m)^2(\varw+2)n^2-6(\gamma-2)(2+m)n-36\gamma+54\big)\\&\bigg(x^{\frac{m}{3}}y^{\frac{1}{3}}\big(\gamma\big(tc_1+c_2\big)\big)^{\frac{(2+m)}{3\gamma}}\bigg)^n+144\pi\gamma\big(\big(3\gamma t^2\xi_{0}+m\xi_{2}+2\xi_{2}\big)c_1^2+6\xi_{0}c_1c_2t\gamma+3\xi_{0}c_2^2\big)\bigg),
	\end{aligned}\right\}
	$$
	$$
	\left.
	\begin{aligned}
	&\varPhi_{13}=\Bigg(\frac{-9}{2}\Bigg(\sinh^{2}\big(k_{1}t+k_{3}\big)\Bigg(\frac{-3 x^{-m} k_{1} \phi_{0} }{8}\bigg(k_{2}^{2}\bigg(k_{1}^{2}+\bigg(\frac{-2 m}{3}-\frac{4}{3}\bigg)k_{1}+\frac{1}{6}(m+2)^{2}(\varw+2)\bigg)(m+2)^{2}\\& \cosh^{4}\big(k_{1}t+k_{3}\big)+\Bigg(\Bigg(\Bigg(\frac{-16}{3}-\frac{11m}{2}-\frac{2 m^{2}}{3}\Bigg)k_{1}^{2}+(m^{2}+2m)k_{1}-\frac{1}{4} \Bigg(\frac{-16}{3}+\big(\varw+\frac{2}{3}\big)m\Bigg)(m+2)^{2}\Bigg)k_{2}^{2}\\&+\frac{k_{1}^{2}(m+4)(m+2) }{3}\Bigg)\cosh^{2}\big(k_{1}t+k_{3}\big)+\bigg(\bigg(\frac{7m}{3}+\frac{5}{3}\bigg)k_{1}^{2}+(2m+4)k_{1}-\frac{1}{2}\big(\varw+\frac{13}{3}\big)(m+2)^{2}\bigg)k_{2}^{2}\\&-\frac{7 k_{1}^{2} (m+2)}{6}\bigg)\cosh\big(k_{1}t+k_{2}\big) \bigg(\frac{k_{2}\sinh\big(k_{1}t+k_{2}\big)}{k_{1}}\bigg)^{-m}+k_{2}^{4}\bigg(\xi_{1} k_{1} (m+2)\cosh^{5}\big(k_{1}t+k_{2}\big)\\&-\frac{2}{3} \sinh\big(k_{1}t+k_{2}\big)\bigg(\xi_{2} (m+2)^{2} k_{1}^{2} +\frac{3 \xi_{0} (m-1)}{2}\bigg)\cosh^{4}\big(k_{1}t+k_{2}\big)-2\xi_{0} k_{1} (m+2)\cosh^{3}\big(k_{1}t+k_{2}\big)\\&+\frac{1}{3} \sinh\big(k_{1}t+k_{2}\big)\bigg(\xi_{2} (m+2)^{2} k_{1}^{2}+\frac{3 \xi_{0} (2m-5)}{2}\bigg) \cosh^{4}\big(k_{1}t+k_{2}\big)+\xi_{1} k_{1} (m+2) \cosh\big(k_{1}t+k_{2}\big)\\&+\frac{1}{2}\bigg(\xi_{2} k_{1}^{2}(m+2)+3\xi_{0}\bigg) \sinh\big(k_{1}t+k_{2}\big)\bigg)\Bigg)\Bigg)\Bigg),
	\end{aligned}\right\}
	$$
	$$
	\left.
	\begin{aligned}
	&\& \quad \varPhi_{14}=\Bigg(3\Bigg(\frac{9}{16}x^{-m}k_{4} \phi_{0}\cosh\big(k_{4}t+k_{6}\big) \bigg(\frac{k_{5}\sinh\big(k_{4}t+k_{6}\big)}{k_{4}}\bigg)^{-m}\Bigg(k_{5}^{2}\Bigg(k_{4}^{2}+\bigg(\frac{-2 m}{3}-\frac{4}{3}\bigg)k_{4}+\frac{1}{6}(m+2)^{2}(\varw+2)\Bigg)\\&(m+2)^{2}\cosh^{4}\big(k_{4}t+k_{6}\big)+\Bigg(\Bigg(\Bigg(\frac{-16}{3}-\frac{11m}{2}-\frac{2 m^{2}}{3}
	\Bigg)k_{4}^{2}+(m^{2}+2m)k_{4}-\frac{1}{4} \Bigg(\frac{-16}{3}+\big(\varw+\frac{2}{3}\big)m\Bigg)(m+2)^{2}\Bigg)k_{5}^{2}\\&-\frac{k_{4}^{2}(m+4)(m+2) }{3}\Bigg)\cosh^{2}\big(k_{4}t+k_{6}\big)+\bigg(\bigg(\frac{7m}{3}+\frac{5}{3}\bigg)k_{4}^{2}+(2m+4)k_{4}-\frac{1}{2}\big(\varw+\frac{13}{3}\big)(m+2)^{2}\bigg)k_{5}^{2}+\frac{7 k_{4}^{2} (m+2)}{6}\Bigg)\\&+\Bigg(\frac{-3}{2} k_{4} \xi_{1} (m+2) \cosh^{5}\big(k_{4}t+k_{6}\big)+\sinh\big(k_{4}t+k_{6}\big)\Bigg(\xi_{2}(m+2)^{2}k_{4}^{2}+\frac{3}{2} \xi_{0} (m-1)\Bigg)\cosh^{4}\big(k_{4}t+k_{6}\big)\\&+3k_{4}\xi_{1}(m+2)\cosh^{3}\big(k_{4}t+k_{6}\big)-\frac{1}{2}\Bigg(\xi_{2} (m+2)^{2} k_{4}^{2}+3\bigg(m-\frac{5}{2}\bigg) \xi_{0}\Bigg)  \sinh\big(k_{4}t+k_{6}\big)\cosh^{2}\big(k_{4}t+k_{6}\big)\\&-\frac{3}{2}\xi_{1} k_{4} (m+2) \cosh\big(k_{4}t+k_{6}\big) -\frac{3}{4}\sinh\big(k_{4}t+k_{6}\big)\bigg(\xi_{2} k_{4}^{2} (m+2)+3\xi_{0}\bigg)\Bigg)k_{5}^{4} \pi \sinh y (m+2)\Bigg)\sinh^{2}\big(k_{4}t+k_{6}\big)\Bigg).
	\end{aligned}\right\}
	$$
	Figure \eqref{fig:eosp01} depict the plots of $\omega_{de}-\omega_{de}'$ plane for all three models with various values of $c_{1}$, $c_{2}$, $k_{1}$, $k_{3}$, $k_{4}$, \& $k_{6}$, respectively. It is observed that for models--I \& II, the $\omega_{de}-\omega_{de}'$ plane is mainly characterized in freezing region ($\omega_{de}'<0$, $\omega_{de}<0$); whereas for model-III, the trajectories vary in both the freezing and thawing region. Ultimately, the trajectories in the above mentioned figures represent the current cosmic expansion in an accelerated manner.
	\begin{figure}[H]
		\centering
		\begin{minipage}{0.32\textwidth}
			\includegraphics[width=1\linewidth]{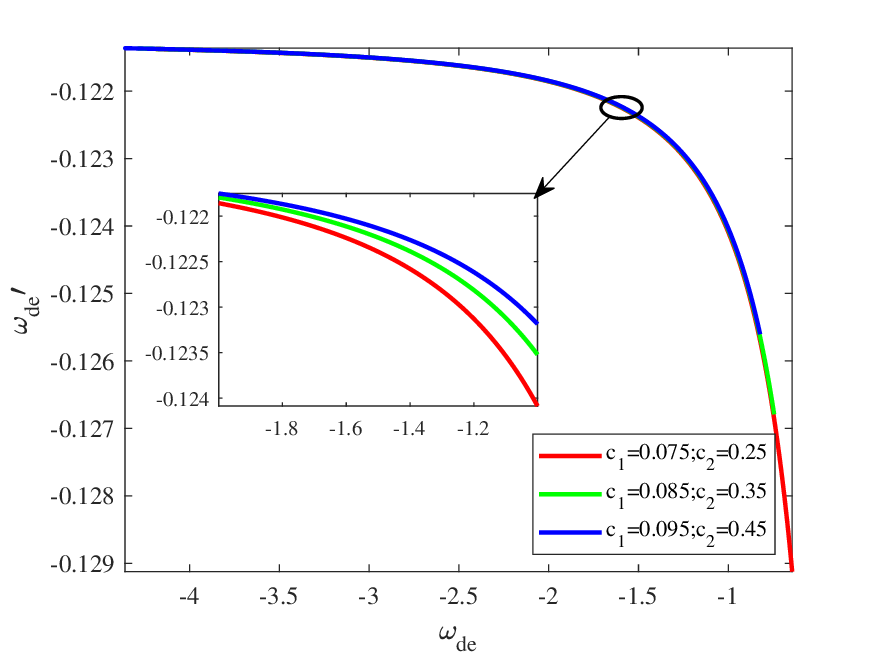}
			\caption*{Model-I}
			\label{fig:eosp0}
		\end{minipage}
		\begin{minipage}{0.32\textwidth}
			\includegraphics[width=1\linewidth]{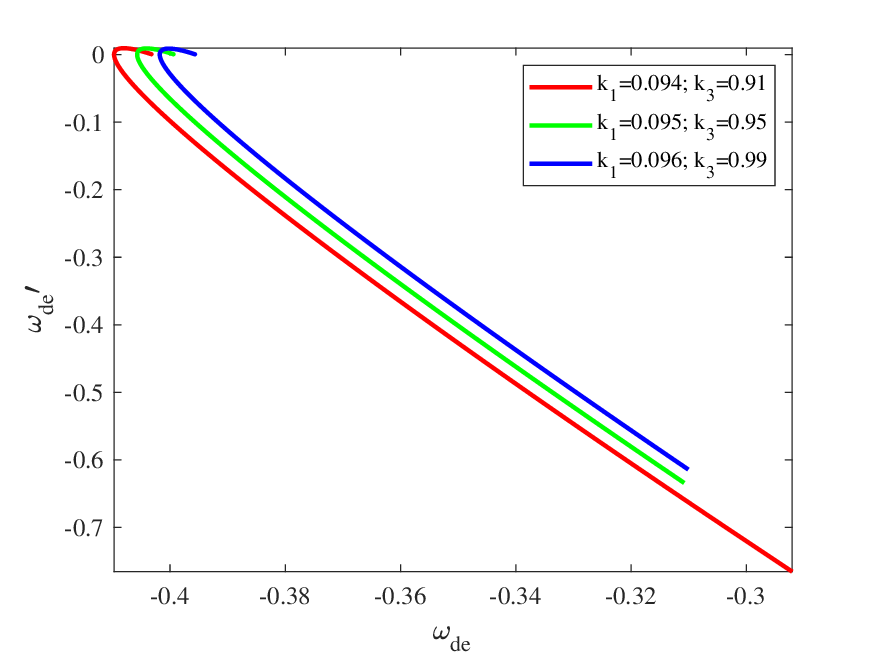}
			\caption*{Model--II}
			\label{fig:eosp1}
		\end{minipage}
		\begin{minipage}{0.32\textwidth}
			\includegraphics[width=1\linewidth]{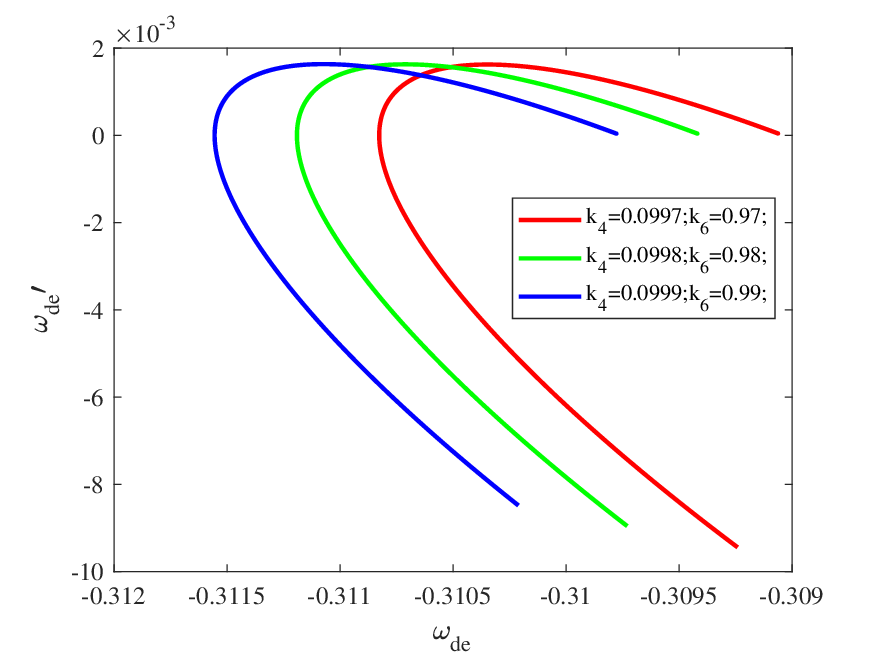}
			\caption*{Model--III}
			\label{fig:eospm1}
		\end{minipage}
		\caption{ Plot of  $\omega_{de}-\omega_{de}'$ plane.}
		\label{fig:eosp01}
	\end{figure}
	\item {\bf Om-diagnostic:} 
	To discriminate among different phases of the Universe viz. the $\Lambda CDM$ for non-minimally coupled scalar field, quintessence model and phantom field, through the trajectories of the curves; a tool introduced by Sahni et al. \cite{vsa}, called as om-diagnostic, plays a vital role. The trajectories of Om-diagnostics determine different eras, such as, a positive trajectory determining phantom DE era, whereas the negative trajectory indicates the quintessence DE. The om-diagnostics is defined as
	\begin{equation}
	\text{Om(z)}= \frac{H^2(z)-H_0^{2}}{H_0^{2}((1+z)^3-1)}.
	\end{equation}
	Hence, for models--I, II \& III we obtained the Om(z) expression as
	\begin{align}
	\text{Om(z)}=&\frac{c_{1}^{2} (m+2)^{2} x^{\frac{2\gamma m}{(m+2)}} y^{\frac{2\gamma}{m+2}} (1+z)^{\frac{6 \gamma}{m+2}} -H_{0}^{2}9 a_{0}^{\frac{6 \gamma}{m+2}}}{H_{0}^{2}9 a_{0}^{\frac{6 \gamma}{m+2}} \bigg((1+z)^{3}-1\bigg)},\\ \text{Om(z)}=&\frac{(m+2)^{2}k_{1}^{2}\bigg(k_{1}^{2}a_{0}^{\frac{6}{m+2}}+ k_{2}^{2} (1+z)^{\frac{6}{m+2}} x^{\frac{2m}{m+2}} (\sin y)^{\frac{6}{m(m+2)}} \bigg)-9k_{1}^{2}a_{0}^{\frac{6}{m+2}} H_{0}^{2} }{9k_{1}^{2}a_{0}^{\frac{6}{m+2}} H_{0}^{2} \big((1+z)^{3}-1\big)},
	\\
	\& \quad \text{Om(z)}=&\frac{(m+2)^{2}k_{4}^{2}\bigg(k_{4}^{2}a_{0}^{\frac{6}{m+2}}+ k_{5}^{2} (1+z)^{\frac{6}{m+2}} x^{\frac{2m}{m+2}} (\sinh y)^{\frac{6}{m(m+2)}} \bigg)-9k_{4}^{2}a_{0}^{\frac{6}{m+2}} H_{0}^{2}}{9k_{4}^{2}a_{0}^{\frac{6}{m+2}} H_{0}^{2} \big((1+z)^{3}-1\big)},
	\end{align}
	\begin{figure}[H]
		\centering
		\begin{minipage}{0.32\textwidth}
			\includegraphics[width=1\linewidth]{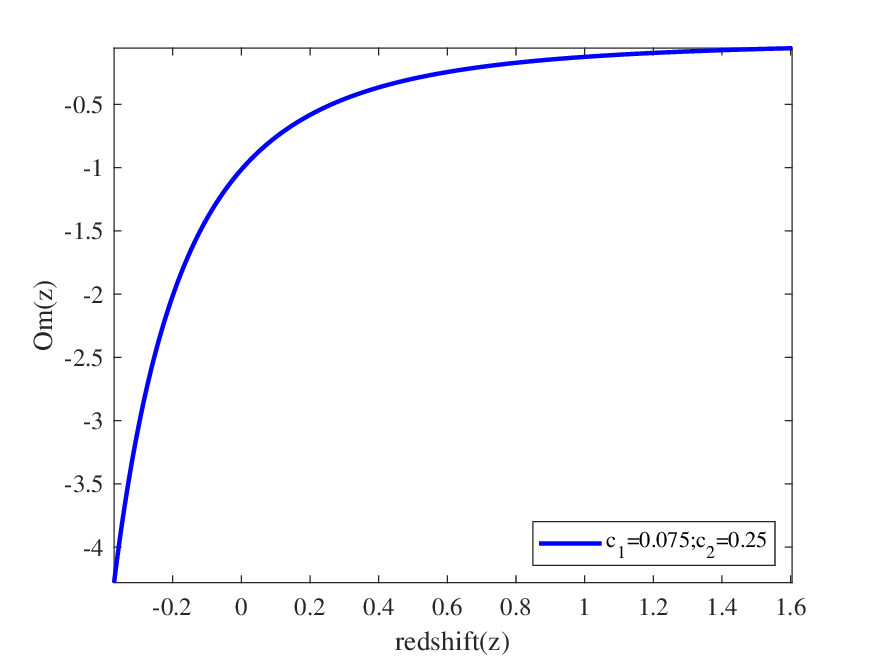}
			\caption*{Model--I}
			\label{fig:omz0}
		\end{minipage}
		\begin{minipage}{0.32\textwidth}
			\includegraphics[width=1\linewidth]{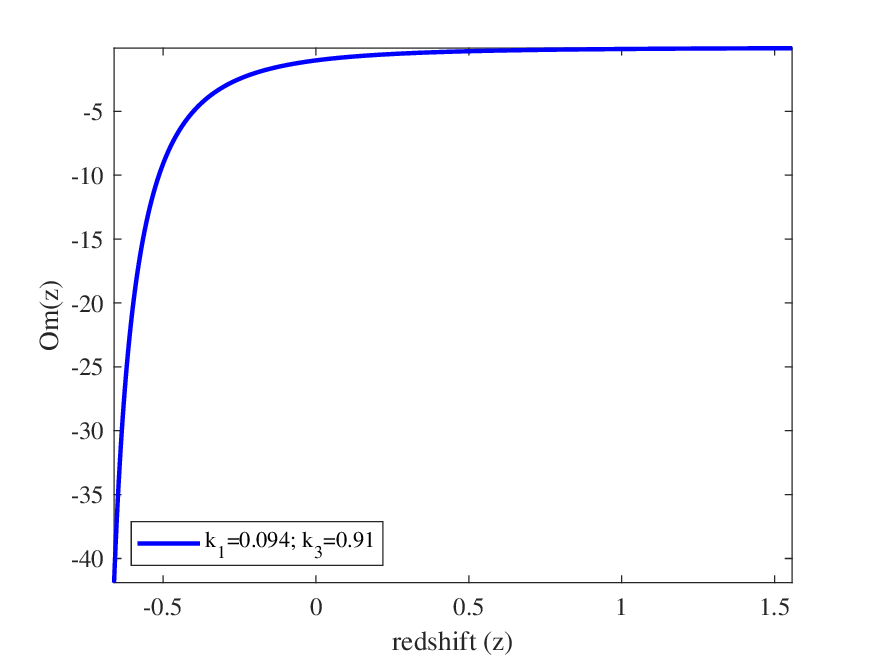}
			\caption*{Model--II}
			\label{fig:omz1}
		\end{minipage}
		\begin{minipage}{0.32\textwidth}
			\includegraphics[width=1\linewidth]{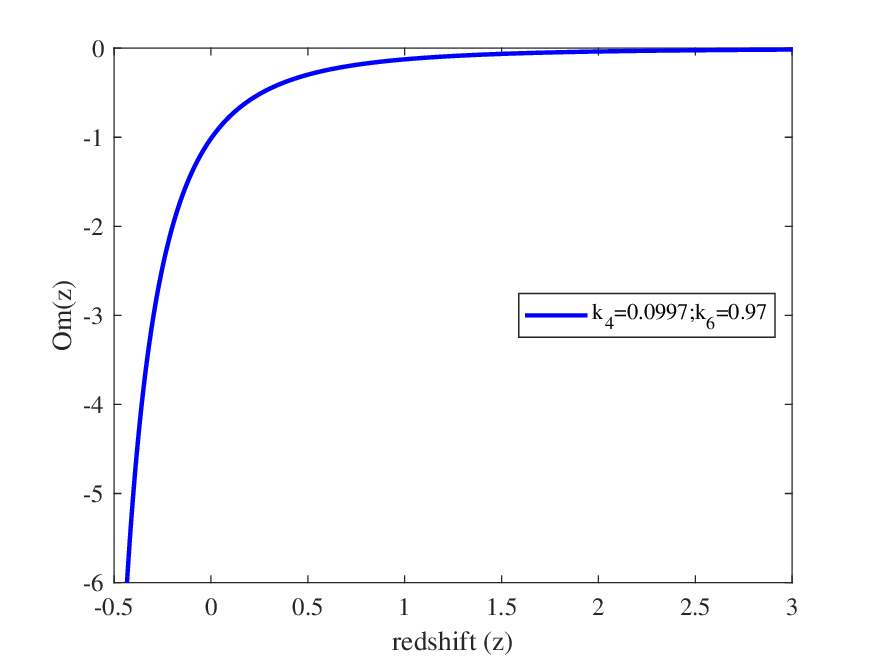}
			\caption*{Model--III}
			\label{fig:omzm1}
		\end{minipage}
		\caption{ Plot of Om(z) versus redshift $(z)$.}
		\label{fig:omz01}
	\end{figure}
	The plot of Om($z$) against redshift ($z$) for all three models are depicted in figure \eqref{fig:omz01}. It can be seen that the trajectories of the parameter in all three figures differ in the negative region, showing the behavior as quintessence DE.
	\item {\bf Stability analysis:} To examine the stability of any DE model, we utilize the squared speed of sound ($v_s^2$). The models with $v_s^2<0$ show unstability where as models with $v_s^2>0$ show stability. Hence, the $v_s^2$ is determined as follows \cite{ysm}:
	\begin{equation}
	v_s^2=\frac{\dot{p_{de}}}{\dot{\rho_{de}}},
	\end{equation} 
	where $\dot{p_{de}}$ and $\dot{\rho_{de}}$ are  the differentiation of pressure and density of DE w.r.t. cosmic time`t' respectively.\\
	The squared speed of the sound for the models--I, II \& III is given by 
	\begin{align}
	\label{eq79}
	v_s^2=&\frac{\varPhi_{15}}{576\big(tc_1+c_2\big)\pi\gamma\big(m-\frac{3}{2}\gamma+2\big)c_1(2+m)\alpha},\\ 
	\label{eq80}
	v_{s}^{2}=&\frac{2\bigg(\varPhi_{16}+\varPhi_{17}\bigg)}{k_{2}^{4} \sinh^{4}\big(k_{1}t+k_{3}\big)\pi \alpha k_{1} (m+2)(2m+1) \sin y\, \cosh\big(k_{1}t+k_{3}\big)},
	\\
	\label{eq81}
	\& \quad v_{s}^{2}=&\frac{2\varPhi_{18}}{k_{4}^{2}k_{5}^{4}\pi \alpha(m+2)(2m+1)\sinh y\,  \sinh^{4}\big(k_{4}t+k_{6}\big) \cosh\big(k_{4}t+k_{6}\big)},
	\end{align}
	$$
\left.
\begin{aligned}
\text{where}\quad &\varPhi_{15}=\bigg(\big(tc_1+c_2\big)((2+m)n-6\gamma)c_1((2+m)^2(w+2)n^2-6(\gamma-2)(2+m)n-36\gamma+54)\\&\phi_0\bigg(x^\frac{m}{3}y^\frac{1}{3}\big(\gamma\big(tc_1+c_2\big)\big)^{\frac{(2+m)}{3\gamma}} \bigg)^n+288\bigg(\bigg(\frac{3}{2}\xi_{0}t^2\gamma+\big(\xi_{1}t-\frac{3}{2}\xi_{2}\big)(2+m)\bigg)c_1^2+c_2(3\xi_{0}t\gamma+\xi_{1}(2\\&+m))c_1+\frac{3}{2}\xi_{0}c_2^2\gamma\bigg) \pi \gamma (m+2)\bigg)\,\,\, ,
\end{aligned}\right\}
$$
$$
\left.
\begin{aligned}
\quad &\varPhi_{16}=\Bigg(\frac{1}{2} \Bigg(\sinh^{2}\big(k_{1}t+k_{2}\big)\Bigg(\frac{-9}{16}\bigg(k_{2}^{2}\bigg(k_{1}^{2}
+\bigg(\frac{-2m}{3}-\frac{-4}{3}\bigg)k_{1}+\frac{1}{6}(m+2)^{2}(\varw+2)\bigg)(2+m)\cosh^{2}\big(k_{1}t+k_{2}\big)\\&+\bigg(\bigg(\frac{-2m}{3}-\frac{-4}{3}\bigg)k_{1}^{2}+\bigg(\frac{-4m}{3}-\frac{-8}{3}\bigg) k_{1}+\frac{1}{3}\bigg((\varw+3)m+2\varw+8\bigg)(m+2)\bigg)k_{2}^{2}+\frac{1}{3}k_{1}^{2}(m+4)\bigg)x^{-m} \phi_{0}k_{1}\\&\cosh\big(k_{1}t+k_{3}\big) \Bigg(\dfrac{k_{2}\sinh\big(k_{1}t+k_{3}\big)}{k_{1}}\Bigg)^{-m}
\Bigg)\Bigg)\Bigg)\,\, ,
\end{aligned}\right\}$$
$$
\left.
\begin{aligned}
\quad &\varPhi_{17}=\Bigg(
+k_{2}^{4}\bigg(\xi_{1} k_{1}(m+2) \cosh^{3}\big(k_{1}t+k_{3}\big)-\sinh\big(k_{1}t+k_{3}\big)\bigg(\xi_{2} k_{1}^{2}(m+2)-\frac{3\xi_{0}}{2}\bigg) \cosh^{2}\big(k_{1}t+k_{3}\big)\\&-\xi_{1} k_{1}(m+2) \cosh\big(k_{1}t+k_{3}\big)-\frac{1}{2}\bigg(\xi_{2} k_{1}^{2}(m+2)+3\xi_{0}\bigg)\sinh\big(k_{1}t+k_{3}\big)\bigg) \pi \sin y (m+2)\Bigg)\,\, ,
\end{aligned}\right\}$$
$$
\left.
\begin{aligned}
\quad &\varPhi_{18}=\Bigg(\frac{-1}{2} \Bigg(\sinh^{2}\big(k_{4}t+k_{6}\big)\Bigg(\frac{9}{16}\bigg(k_{5}^{2}\bigg(k_{4}^{2}
+\bigg(\frac{-2m}{3}-\frac{-4}{3}\bigg)k_{4}+\frac{1}{6}(m+2)^{2}(\varw+2)\bigg)(2+m)\cosh^{2}\big(k_{4}t+k_{6}\big)\\&+\bigg(\bigg(\frac{-2m}{3}-\frac{-4}{3}\bigg)k_{4}^{2}+\bigg(\frac{-4m}{3}-\frac{-8}{3}\bigg) k_{4}+\frac{1}{3}\bigg((\varw+3)m+2\varw+8\bigg)(m+2)\bigg)k_{5}^{2}+\frac{1}{3}k_{4}^{2}(m+4)\bigg)\cosh\big(k_{4}t+k_{6}\big)\\&x^{-m} \phi_{0}k_{4} \Bigg(\frac{k_{5}\sinh\big(k_{4}t+k_{6}\big)}{k_{4}}\Bigg)^{-m}+k_{5}^{4}\bigg(-\xi_{1} k_{4}(m+2) \cosh^{3}\big(k_{4}t+k_{6}\big)-\sinh\big(k_{4}t+k_{6}\big)\bigg(\xi_{2} k_{4}^{2}(m+2)-\frac{3\xi_{0}}{2}\bigg)\\& \cosh^{2}\big(k_{4}t+k_{6}\big)-\xi_{1} k_{4}(m+2) \cosh\big(k_{4}t+k_{6}\big)-\frac{1}{2}\bigg(\xi_{2} k_{4}^{2}(m+2)+3\xi_{0}\bigg)\sinh\big(k_{4}t+k_{6}\big)\bigg) \pi \sinh y (m+2)\Bigg)\Bigg)\Bigg).
\end{aligned}\right\}
$$\vspace{-1pc}
	\begin{figure}[H]
		\centering
		\begin{minipage}{0.32\textwidth}
			\includegraphics[width=1\linewidth]{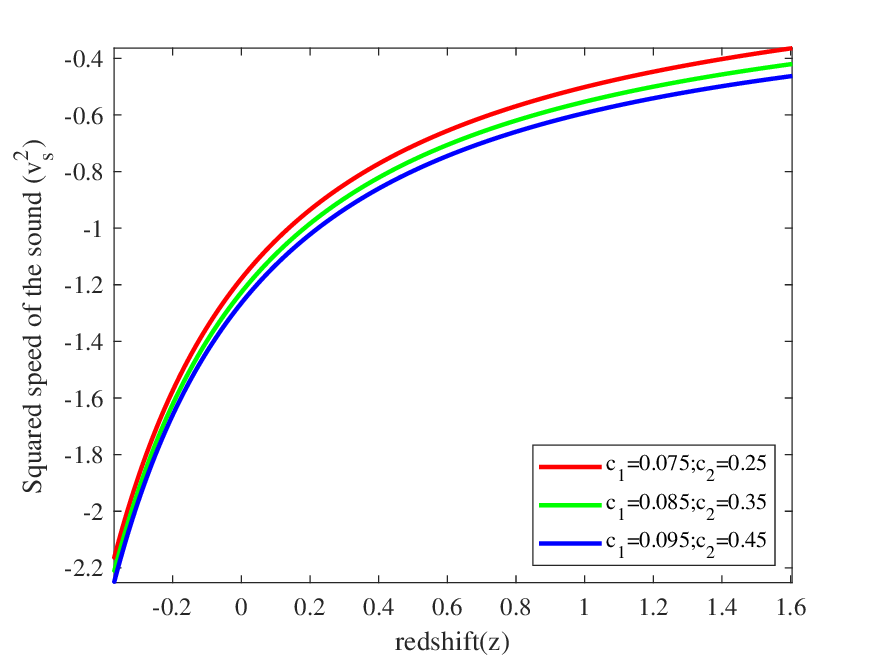}
			\caption*{Model--I}
			\label{fig:vs0}
		\end{minipage}
		\begin{minipage}{0.32\textwidth}
			\includegraphics[width=1\linewidth]{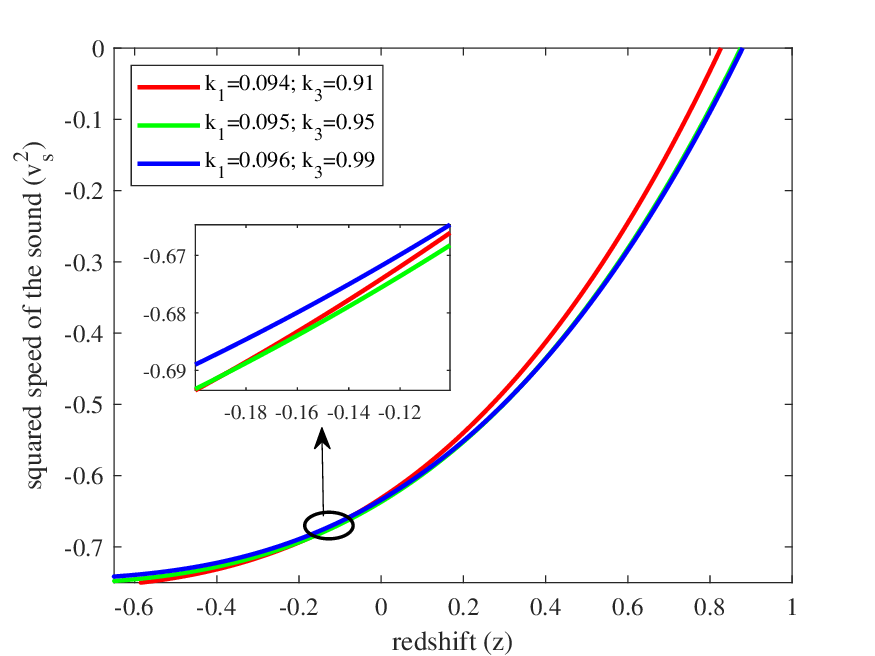}
			\caption*{Model--II}
			\label{fig:vs1}
		\end{minipage}
		\begin{minipage}{0.32\textwidth}
			\includegraphics[width=1\linewidth]{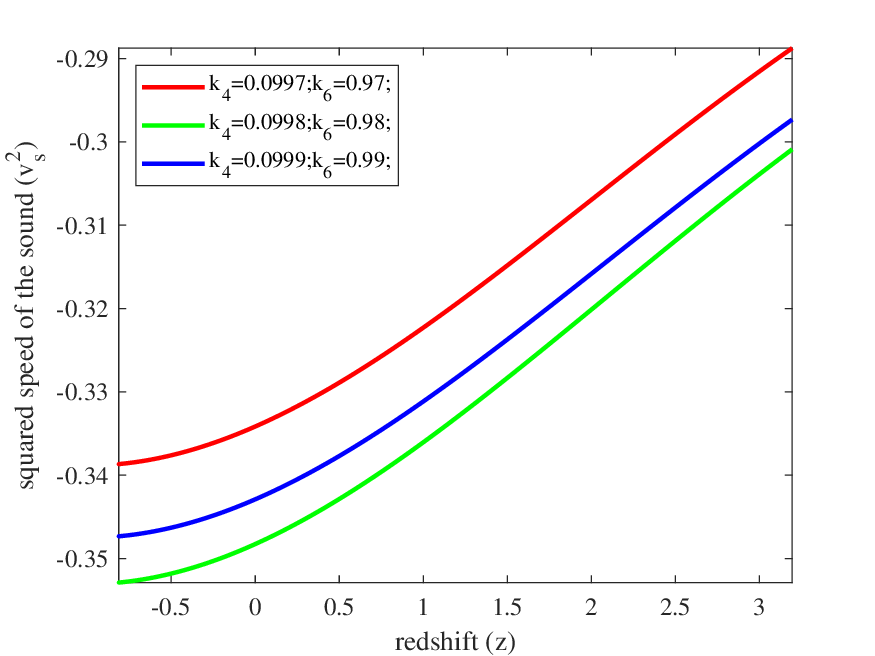}
			\caption*{Model--III}
			\label{fig:vsm1}
		\end{minipage}
		\caption{ Plot of squared speed of the sound versus redshift $(z)$.}
		\label{fig:vs01}
	\end{figure}
	To determine the stability of the obtained models--I, II  \& III, the squared speed of sound ($v_s^2$) against redshift($z$) is plotted in figure \eqref{fig:vs01} for the values of $c_{1}$, $c_{2}$, $k_{1}$, $k_{3}$, $k_{4}$, \& $k_{6}$, respectively. The models represent an unstable behavior of the Universe as the trajectories are varying in the negative region. 
\end{itemize}
\section{Conclusions}\label{c8con}
In this paper, we have analyzed the field equations of BDT for VRDE in Ruban's space-time. After the evolution of the field equations to acquire the scale factor and other cosmological parameters, three possibilities are identified for $\kappa$ as $0$, $1$ and $-1$. For all three obtained models, we have plotted the trajectories for various parameters against redshift($z$) to scrutinize the behavior of the cosmos.\\
The constructed models which are mentioned in Eqs. \eqref{eq21}, \eqref{eq34} and \eqref{eq46} are anisotropic and expanding as $t\rightarrow \infty$. The DP ($q$) of the models for $\kappa$=1 \& -1 traverse from the decelerated phase of the past to the accelerated phase of the present. The jerk parameter of the models with $\kappa$=1 and -1 traverse in the positive region and $j\rightarrow1$ in near future as $z\rightarrow0$. However, for model-I, the DP and the jerk parameter are independent of time. For models--II \& III, the statefinder pair has $\Lambda$CDM region and also has quintessence and phantom regions in their transition, whereas for model--I, the pair $(r,s)$ is independent of time. The trajectories of $q-r$ plane shows a signature change from negative region to positive region ($i.e.,$ the trajectories are traveling from radiation dominated era to matter dominated region) and finally reach the de-sitter phase of the Universe. The EoS parameter for $\kappa$=0, the Universe shows quintom-like behavior, as the path of $\omega_{de}$ travel from quintessence to phantom region, and for $\kappa=1$ \& $-1$, it represents the quintessence nature by completely varying in quintessence region. The $\omega_{de}-\omega_{de}'$ plane for the models--I \& II has mainly characterized in freezing region and for model-III the trajectories differ in both freezing and thawing regions. Eventually, by the analysis of $\omega_{de}-\omega_{de}'$ plane we can conclude that the expansion of the cosmos, in the present times, is in an accelerated manner. The trajectories of the square speed of sound ($v_s^2$) vary in the negative region depicting an unstable behavior of the Universe. And finally the study of Om($z$) diagnostics says that the DE models represent the quintessence behavior, as the three models have the negative values of Om(z). Therefore, our models show an anisotropic behavior with the accelerated expansion phenomenon, justifying the ongoing research around the globe. 
\section*{Acknowledgments}
MVS acknowledges Department of Science and Technology (DST), Govt of India, New Delhi for financial support to carry out the Research Project [No. EEQ/2021/000737, Dt. 07/03/2022]. The authors are very much thankful to the editorial team and the reviewer’s for their constructive comments and valuable suggestions which have certainly improved the presentation and quality of the paper.

\end{document}